\documentclass[longauth]{aa} 
\usepackage{graphicx}
\usepackage{txfonts}
\usepackage{amsmath,amsfonts,amssymb} 
\usepackage{natbib}
\usepackage{multicol,caption}
\usepackage{lipsum}
\usepackage{booktabs}
\usepackage{threeparttable}
\usepackage{multirow}
\usepackage{here}
\usepackage{lscape}

\bibpunct{(}{)}{;}{a}{}{,}
\usepackage[colorlinks=true,urlcolor=blue,linkcolor=blue,citecolor=blue]{hyperref}

\newcommand{\micron}{$\mu$m}

\newcommand{\tls}{{\fontfamily{pcr}\selectfont tls}\,}
\newcommand{\allesfitter}{{\fontfamily{pcr}\selectfont allesfitter}\,}
\newcommand{\sherlock}{{\fontfamily{pcr}\selectfont SHERLOCK}\,}
\newcommand{\triceratops}{{\fontfamily{pcr}\selectfont TRICERATOPS}\,}
\newcommand{\lightkurve}{{\fontfamily{pcr}\selectfont lightkurve}\,}

\usepackage{threeparttable}
\usepackage{booktabs}
\usepackage{longtable}
\usepackage{ragged2e}

\usepackage{tikz,xcolor,hyperref}

\definecolor{lime}{HTML}{A6CE39}
\DeclareRobustCommand{\orcidicon}{%
	\hspace{-1.5mm}
	\begin{tikzpicture}
	\draw[lime, fill=lime] (0,0) 
	circle [radius=0.16] 
	node[white] {{\fontfamily{qag}\selectfont \tiny ID}};
	\draw[white, fill=white] (-0.0625,0.095) 
	circle [radius=0.007];
	\end{tikzpicture}
	\hspace{-2.5mm}
}
\foreach \x in {A, ..., Z}{%
	\expandafter\xdef\csname orcid\x\endcsname{\noexpand\href{https://orcid.org/\csname orcidauthor\x\endcsname}{\noexpand\orcidicon}}
}

\foreach \x in {a, ..., z}{%
	\expandafter\xdef\csname orcid\x\endcsname{\noexpand\href{https://orcid.org/\csname orcidauthor\x\endcsname}{\noexpand\orcidicon}}
}

\begin{document}

\title{TOI-2084\,b and TOI-4184\,b: two new sub-Neptunes around M dwarf stars}

\author{
	K. Barkaoui\orcidA{}\inst{\ref{astro_liege},\ref{MIT},\ref{IAC_Laguna}} \thanks{E-mail: \color{blue}khalid.barkaoui@uliege.be} 
	\and M. Timmermans\inst{\ref{astro_liege}}  
	\and A. Soubkiou\inst{\ref{ouka},\ref{porto-fis},\ref{porto-iace}}
	\and B.V.~Rackham\orcidC{}\inst{\ref{MIT},\ref{Kavli_MIT}} 
	\and A.~J.~Burgasser\orcidD{}\inst{\ref{UCSDiego}} 
    \and J.~Chouqar\inst{\ref{ouka}} 
    \and F.J.~Pozuelos\orcidE{}\inst{\ref{iaa}} 
    \and K.A. Collins\orcidF{}\inst{\ref{Harvard_USA}} 
	\and S.B.~Howell\orcidG{}\inst{\ref{Ames_NASA}} 
	\and R.~Simcoe\inst{\ref{UCSDiego}} 
	\and C.~Melis\orcidH{}\inst{\ref{UCSDiego}} 
	\and K.G.~Stassun\orcidI{}\inst{\ref{vanderbilt}} 
	\and J.~Tregloan-Reed\inst{\ref{Inst_Chili_M4}} 
	\and M.~Cointepas\inst{\ref{Alpes_Grenoble},\ref{Dep_Astro_Geneve}} 
	\and M.~Gillon\orcidJ{}\inst{\ref{astro_liege}} 
	\and X.~Bonfils\orcidK{}\inst{\ref{Alpes_Grenoble}} 
	\and E.~Furlan\orcidL{}\inst{\ref{Exoplanet_NASA}} 
	\and C.L.~Gnilka\inst{\ref{Ames_NASA}} 
    \and J.M.~Almenara\orcidp{}\inst{\ref{Alpes_Grenoble}} 
    \and R.~Alonso\orcidw{}\inst{\ref{IAC_Laguna},\ref{la_laguna}} 
    \and Z.~Benkhaldoun\orcidl{}\inst{\ref{ouka}} 
    \and M.~Bonavita\orcidq{}\inst{\ref{Edi_Uni_M5}} 
    \and F.~Bouchy\inst{\ref{Dep_Astro_Geneve}} 
    \and A.~Burdanov\orcidN{}\inst{\ref{MIT}} 
    \and P.~Chinchilla\inst{\ref{IAC_Laguna}} 
    \and F.~Davoudi\orcidm{}\inst{\ref{astro_liege}} 
    \and L.~Delrez\orcidO{}\inst{{\ref{astro_liege},\ref{star_liege}}} 
    \and O.~Demangeon\inst{\ref{porto-fis},\ref{porto-iace}} 
    \and M.~Dominik\orcidv{}\inst{\ref{Exop_Sc_M2}}  
    \and B.-O.~Demory\orcidJ{}\inst{\ref{unibe}} 
    \and J.~de~Wit\inst{\ref{MIT}} 
    \and G.~Dransfield\orcidQ{}\inst{\ref{birmingham}} 
    \and E.~Ducrot\orcidR{}\inst{\ref{Paris_Region},\ref{cea}} 
    \and A.~Fukui\orcidS{}\inst{\ref{Univ_tokyo},\ref{IAC_Laguna}},
    \and T.~C.~Hinse\inst{\ref{Cop_Univ_M6}} 
    \and M.J.~Hooton\orcidT{}\inst{\ref{Cavendish}} 
    \and E.~Jehin\orcidU{}\inst{\ref{star_liege}} 
    \and J.~M.~Jenkins\orcidV{}\inst{\ref{Ames_NASA}} 
    \and U.~G.~J{\o}rgensen\orcids{}\inst{\ref{ExoLife_M1}}   
    \and D.~W.~Latham\orcidW{}\inst{\ref{Harvard_USA}} 
    \and L.~Garcia\orcidY{}\inst{\ref{astro_liege}} 
    \and S.~Carrazco-Gaxiola\inst{\ref{ciudad},\ref{gsu},\ref{recons}} 
    \and M.~Ghachoui\inst{\ref{ouka},\ref{astro_liege}} 
    \and Y.~G\'omez~Maqueo~Chew\orcidM{}\inst{\ref{ciudad}} 
    \and M.N.~G\"unther\orcidr{}\inst{\ref{estec}} 
    \and J.~McCormac\inst{\ref{warwick}} 
    \and F.~Murgas\orcidX{}\inst{\ref{IAC_Laguna},\ref{la_laguna}} 
    \and C.~A.~Murray\orcidZ{}\inst{\ref{Colorado}} 
    \and N.~Narita\orcida{}\inst{\ref{Univ_tokyo},\ref{Astro_tokyo},\ref{IAC_Laguna}} 
    \and P.~Niraula\orcidb{}\inst{\ref{MIT}} 
    \and P.~P.~Pedersen\orcidj{}\inst{\ref{Cavendish}} 
    \and D.~Queloz\orcidc{}\inst{\ref{Cavendish}} 
    \and R.~Rebolo-L\'opez\orcidd{}\inst{\ref{IAC_Laguna},\ref{la_laguna}} 
    \and G.~Ricker\inst{\ref{Kavli_MIT}} 
    \and L.~Sabin\orcidt{}\inst{\ref{uname}} 
    \and S.~Sajadian\orcido{}\inst{\ref{Isfahan_M7}} 
    \and N.~Schanche\inst{\ref{unibe}} 
    \and R.~P.~Schwarz\orcidi{}\inst{\ref{Harvard_USA}} 
    \and S.~Seager\inst{\ref{UCSDiego},\ref{la_laguna},\ref{Univ_of_Maryl}} 
    \and D.~Sebastian\orcide{}\inst{\ref{birmingham}} 
    \and R.~Sefako\orcidu{}\inst{\ref{saao}} 
    \and S.~Sohy\inst{\ref{star_liege}} 
    \and J.~Southworth\orcidn{}\inst{\ref{Keele_Uni_M3}} 
    \and G.~Srdoc\inst{\ref{Kotiza_Obser}} 
    \and S.~J.~Thompson\inst{\ref{Cavendish}} 
    \and A.~H.~M.~J.~Triaud\orcidf{}\inst{\ref{birmingham}} 
    \and R.~Vanderspek\inst{\ref{Kavli_MIT}}  
    \and R.~D.~Wells\orcidh{}\inst{\ref{unibe}} 
    \and J.~N.~Winn\inst{\ref{Astro_Prin}}   
    \and S.~Z\'u\~niga-Fern\'andez\orcidB{}\inst{\ref{astro_liege}} 
	}

\institute{
	Astrobiology Research Unit, Universit\'e de Li\`ege, All\'ee du 6 Ao\^ut 19C, B-4000 Li\`ege, Belgium \label{astro_liege}
	\and Department of Earth, Atmospheric and Planetary Science, Massachusetts Institute of Technology, 77 Massachusetts Avenue, Cambridge, MA 02139, USA \label{MIT}
	\and Instituto de Astrof\'isica de Canarias (IAC), Calle V\'ia L\'actea s/n, 38200, La Laguna, Tenerife, Spain \label{IAC_Laguna}
	\and Oukaimeden Observatory, High Energy Physics and Astrophysics Laboratory, Faculty of sciences Semlalia, Cadi Ayyad University, Marrakech, Morocco \label{ouka}
    \and Departamento de Fisica e Astronomia, Faculdade de Ciencias,  Universidade do Porto, Rua do Campo Alegre, 4169-007 porto, Portugal \label{porto-fis}
	\and Instituto de Astrofisica e Ciencias do Espaco, Universidade do porto, CAUP, Rua das Estrelas, 150-762 Porto, Portugal \label{porto-iace}
    \and Department of Physics and Kavli Institute for Astrophysics and Space Research, Massachusetts Institute of Technology, Cambridge, MA 02139, USA \label{Kavli_MIT}
    \and Center for Astrophysics and Space Sciences, UC San Diego, UCSD Mail Code 0424, 9500 Gilman Drive, La Jolla, CA 92093-0424, USA \label{UCSDiego}
    \and Instituto de Astrof\'isica de Andaluc\'ia (IAA-CSIC), Glorieta de la Astronom\'ia s/n, 18008 Granada, Spain \label{iaa}
	\and Center for Astrophysics \textbar  Harvard \& Smithsonian, 60 Garden St, Cambridge, MA 02138, USA \label{Harvard_USA}
	\and NASA Ames Research Center, Moffett Field, CA 94035, USA \label{Ames_NASA}
    \and Department of Physics \& Astronomy, Vanderbilt University, 6301 Stevenson Center Ln., Nashville, TN 37235, USA \label{vanderbilt}
    \and Instituto de Astronomía y Ciencias Planetarias de Atacama,  Universidad de Atacama, Copayapu 485,  Copiapó, Chile \label{Inst_Chili_M4}
    \and Univ. Grenoble Alpes, CNRS, IPAG, F-38000 Grenoble, France \label{Alpes_Grenoble}
    \and Observatoire de Genève, Département d'Astronomie, Université de Genève, Chemin Pegasi 51, 1290 Versoix, Switzerland \label{Dep_Astro_Geneve}
	\and NASA Exoplanet Science Institute, Caltech/IPAC, Mail Code 100-22, 1200 E. California Blvd., Pasadena, CA 91125, USA \label{Exoplanet_NASA}
    \and Departamento de Astrof\'{i}sica, Universidad de La Laguna (ULL), 38206 La Laguna, Tenerife, Spain \label{la_laguna}
    \and Institute for Astronomy, The University of Edinburgh, Royal Observatory, Blackford Hill, Edinburgh EH9 3HJ, UK \label{Edi_Uni_M5} 
	\and Space Sciences, Technologies and Astrophysics Research (STAR) Institute, Universit\'e de Li\`ege, All\'ee du 6 Ao\^ut 19C, B-4000 Li\`ege, Belgium \label{star_liege}
    \and University of St Andrews, Centre for Exoplanet Science, SUPA School of Physics \& Astronomy, North Haugh, St Andrews, KY16 9SS, United Kingdom \label{Exop_Sc_M2}
    \and Center for Space and Habitability, University of Bern, Gesellschaftsstrasse 6, 3012, Bern, Switzerland \label{unibe}
    \and School of Physics \& Astronomy, University of Birmingham, Edgbaston, Birmingham B15 2TT, UK \label{birmingham}
    \and Paris Region Fellow, Marie Sklodowska-Curie Action \label{Paris_Region}
    \and AIM, CEA, CNRS, Universit\'e Paris-Saclay, Universit\'e de Paris, F-91191 Gif-sur-Yvette, France \label{cea}
    \and Komaba Institute for Science, The University of Tokyo, 3-8-1 Komaba, Meguro, Tokyo 153-8902, Japan \label{Univ_tokyo}
    \and University of Southern Denmark, University Library, Campusvej 55, DK 5230 Odense M, Denmark \label{Cop_Univ_M6}
    \and Cavendish Laboratory, JJ Thomson Avenue, Cambridge CB3 0HE, UK \label{Cavendish}
    \and Centre for ExoLife Sciences, Niels Bohr Institute, University of Copenhagen, {\O}ster Voldgade 5, 1350 Copenhagen, Denmark \label{ExoLife_M1}
    \and Universidad Nacional Aut\'onoma de M\'exico, Instituto de Astronom\'ia, AP 70-264, Ciudad de M\'exico,  04510, M\'exico \label{ciudad}
    \and European Space Agency (ESA), European Space Research and Technology Centre (ESTEC), Keplerlaan 1, 2201 AZ Noordwijk, The Netherlands \label{estec}
    \and Department of Physics and Astronomy, Georgia State University, Atlanta, GA 30302-4106, USA \label{gsu}
    \and RECONS Institute, Chambersburg, PA 17201, USA \label{recons}
    \and Department of Physics, University of Warwick, Gibbet Hill Road, Coventry CV4 7AL, United Kingdom \label{warwick} 
    \and Department of Astrophysical and Planetary Sciences, University of Colorado Boulder, Boulder, CO 80309, USA \label{Colorado}
    \and Astrobiology Center, 2-21-1 Osawa, Mitaka, Tokyo 181-8588, Japan \label{Astro_tokyo}
    \and Universidad Nacional Aut\'onoma de M\'exico, Instituto de Astronom\'ia, AP 106, Ensenada 22800, BC, M\'exico \label{uname}
    \and Department of Physics, Isfahan University of Technology, Isfahan, Iran \label{Isfahan_M7}
    \and Department of Astronomy, University of Maryland, College Park, MD 20742, USA \label{Univ_of_Maryl}
    \and South African Astronomical Observatory, P.O. Box 9, Observatory, Cape Town 7935, South Africa \label{saao}
    \and Astrophysics Group, Keele University, Staffordshire, ST5 5BG, UK \label{Keele_Uni_M3} 
	\and Kotizarovci Observatory, Sarsoni 90, 51216 Viskovo, Croatia \label{Kotiza_Obser}
    \and Department of Astrophysical Sciences, Princeton University, Princeton, NJ 08544, USA \label{Astro_Prin}
}

\date{Received/accepted}
\titlerunning{TOI-2084\,b \& TOI-4184\,b}\authorrunning{K. Barkaoui et al.}	
	
\abstract{
	We present the discovery and validation of two \emph{TESS} exoplanets orbiting nearby M dwarfs: TOI-2084\,b, and TOI-4184\,b.
	We characterized the host stars by combining spectra from Shane/Kast and Magellan/FIRE, SED (Spectral Energy Distribution) analysis, and stellar evolutionary models.
	In addition, we used Gemini-South/Zorro \& -North/Alopeke high-resolution imaging, archival science images, and statistical validation packages to support the planetary interpretation.
	We performed a global analysis of multi-colour photometric data from \emph{TESS} and ground-based facilities in order to derive the stellar and planetary physical parameters for each system.
	We find that TOI-2084\,b and TOI-4184\,b are sub-Neptune-sized planets with radii of $R_p = 2.47 \pm 0.13R_\oplus$ and $R_p = 2.43 \pm 0.21R_\oplus$, respectively.
	 TOI-2084\,b completes an orbit around its host star every 6.08~days, has an equilibrium temperature of $T_{\rm eq} = 527 \pm 8~K$ and an irradiation of $S_p = 12.8 \pm 0.8~S_\oplus$. Its host star is a dwarf of spectral M$2.0\pm 0.5$ at a distance of 114~pc with an effective temperature of $T_{\rm eff} = 3550\pm 50~K$, and has a wide, co-moving M8 companion at a projected separation of 1400~au.
 
   TOI-4184\,b orbits around an M$5.0\pm 0.5$ type dwarf star ($K_{\rm mag} = 11.87$) each 4.9\,days, and has an equilibrium temperature of $T_{\rm eq} = 412 \pm 8~K$ and an irradiation of $S_p = 4.8 \pm 0.4~S_\oplus$.   TOI-4184 is a metal poor star ($\mathrm{[Fe/H]} = -0.27 \pm 0.09$~dex) at a distance of 69~pc with an effective temperature of $T_{\rm eff} = 3225 \pm 75~K$. 
   Both planets are located at the edge of the sub-Jovian desert in the radius-period plane. The combination of the small size and the large infrared brightness of their host stars make these new planets promising targets for future atmospheric exploration with \emph{JWST}.
}


\maketitle

\section{Introduction}

M dwarfs are the most common stars in our  galaxy \citep{henry1994solar,Kirkpatrick_1999ApJ}, and  small planets occur around M dwarfs more frequently than Sun-like stars \citep{Nutzman_2008,Kaltenegger_2009,Winters_2015}.
M dwarfs are, therefore,  attractive and exciting targets for searching for small and temperate exoplanets using the transit technique, thanks to their small sizes, low masses, and luminosities. The transit signal is much deeper than that caused by similar planets orbiting Sun-like stars, which makes such planets easier to detect and characterize. Moreover, such planetary systems are suitable targets for atmospheric characterization through transmission spectroscopy, including with \emph{JWST} \citep{kem}. In addition, the radial-velocity semi-amplitudes of the stellar hosts are higher, thanks to the low stellar masses, which makes them suitable targets for planetary mass measurements. 
 M dwarf systems will allow a better understanding of the so-called radius valley between the super-Earth- and sub-Neptune-sized planets (see, e.g., \cite{Owen_2013,Fulton_2018,Eylen_2018MNRAS}. Moreover, the discovery of additional sub-Neptune desert planets \citep{Mazeh_2016AandA_Jovian_desert} allows us to further explore and understand the physical properties of such exoplanetary systems.

The Transiting Exoplanet Survey Satellite (\emph{TESS}) mission \citep{Ricker_2015JATIS_TESS} was launched by NASA in 2018 to search for planets around bright nearby dwarfs, including M-type stars.
To date, \emph{TESS} has discovered more than 330 exoplanets orbiting FGKM stars, including 66 planets orbiting around M dwarfs (\href{https://exoplanetarchive.ipac.caltech.edu/}{\it NASA Archive of Exoplanets}).
Since 2018 NASA's \emph{TESS} mission has discovered several sub-Neptune-sized exoplanets around M dwarfs 
(e.g.,
TOI-1696\,b \& TOI-2136\,b: \cite{Beard_2022AJ_TOI1696_TOI2136}
TOI-1201\,b: \cite{Kossakowski_2021A&A_TOI1201b},
TOI-2081\,b \& TOI-4479\,b: \cite{Esparza_2022A&A_TOI4479_TOI2081},
TOI-122\,b \& TOI-237\,b: \cite{Waalkes_2021AJ_TOI122b_TOI237b},
TOI-269\,b: \cite{Cointepas_2021A&A_TOI269b},
TOI-2406\,b: \cite{Wells_2021A&A_TOI2406b},
TOI-620\,b: \cite{Reefe_2022AJ_TOI620b},
TOI-2136\,b: \cite{Tianjun_2022MNRAS.514.4120G},
TOI-2257\,b: \cite{Schanche_2022A&A} and
TOI-2096\,c: \cite{Pozuelos_2023A&A}
).
In this paper we present the discovery and validation of two  new \emph{TESS} exoplanets orbiting nearby M dwarfs, TOI-2084\,b and TOI-4184\,b.  In Section~\ref{photometric_observation}, we present the \emph{TESS} photometry,  high-precision photometric follow-up observations using ground-based facilities, and high-resolution imaging from \textit{Gemini}. 
In Section \ref{stellar_carac}, we present an analysis of the host star properties derived from their Spectral Energy Distributions (SEDs) and spectra.
In Section \ref{Validate_planet}, we validate the planetary nature of the transit signals.
In Section~\ref{Data_analys}, we present our global analysis of the photometric data sets of the planetary systems, which allow us to determine the physical parameters of the star and planet. In Section \ref{search}, we present planet searches and detection limits from the TESS photometry.
Finally, we discuss our results and present our conclusions in Section~\ref{Result_discuss}.

\section{Observation and data reduction} \label{photometric_observation}

\subsection{\emph{TESS} photometry}

The host star TIC\,394357918 (TOI-4184) was observed by \emph{TESS}, \citep{Ricker_2015JATIS_TESS} mission   in Sectors 1,  28 and 39 for 27 days each on \emph{TESS} CCD 3 Camera 3. The Sector 1 campaign started on UTC July 25 2018  and ended on UTC August 22 2018. The Sector 28 campaign started on UTC July 30 2020 and ended on UTC August 26 2020. The Sector 39 campaign started on UTC 2021 May 26 and ended on UTC 2021 June 26. 

 The star TIC\,441738827 (TOI-2084) was observed by \emph{TESS} in 2-minutes cadence during Sectors 16 (UTC September 11 to October 07 2019), 19--23 (UTC November 27 2019 to April 16 2020),  25--26 (UTC May 13 to July 04 2020), 48--60 (UTC January 31 2021 to January 18 2023). TOI-4184 and TOI-2084 were selected by \cite{Stassun_2018AJ_TESS_Catalog} to be observed using the 2-minute short-cadence mode. 
To perform \emph{TESS} data modeling, we retrieved the Presearch Data Conditioning light curves (PDC-SAP, \cite{Stumpe_2012PASP,Smith_2012PASP,Stumpe_2014} constructed by the TESS Science Processing Operations Center (SPOC; \cite{SPOC_Jenkins_2016SPIE}) at Ames Research Center from the Mikulski Archive for Space Telescopes.
PDC-SAP light curves have been corrected for instrument systematics and crowding effects. \autoref{Target_pixel} shows the \emph{TESS} field-of-view for each target and photometric apertures used with the location of nearby Gaia DR3 sources around each target \citep{Gaia_Collaboration_2021A&A}.
\emph{TESS} light curves for TOI-2084 and TOI-4184 are presented in  \autoref{plot_lightcurves_TESS} and \autoref{TOI_2084_4184_TESS_LC}.

\subsection{Ground-based photometry}

We used the {\tt TESS Transit Finder} tool, which is a  customized version of the {\tt Tapir} software package \citep{jensen2013}, to schedule the photometric time-series follow-up observations. These are summarized in the following, and the resulting light curves are presented in \autoref{plot_lightcurves_GB}.

\subsubsection{SPECULOOS-South} \label{sso}

We used one of the SPECULOOS-South (\href{https://www.speculoos.uliege.be}{Search for habitable Planets EClipsing ULtra-cOOl Stars}, \cite{Jehin2018Msngr,Delrez2018,Sebastian_2021AA}  facilities to observe one full transit of TOI-4184.01 on UTC September 25 2021 in the Sloan-$z'$ filter with an exposure time of 42s. Each 1.0-m robotic telescope is equipped with a 2K$\times$2K CCD camera with a pixel scale of 0.35$\arcsec$ and a field of view of 12$\arcmin$$\times$12$\arcmin$.
 We performed aperture photometry in an uncontaminated target aperture of 3.9\arcsec and a PSF full-width half-maximum ({\it FWHM}) of 1.7\arcsec.
Data reduction and  photometric measurements were performed using the {\tt PROSE}\footnote{{\tt Prose:} \url{https://github.com/lgrcia/prose}} pipeline \citep{garcia2021}.

\subsubsection{SPECULOOS-North}

We used SPECULOOS-North/Artemis to observe two transits of TOI-2084.01. Artemis is a 1.0-m Ritchey-Chretien telescope equipped with  a thermoelectrically cooled 2K$\times$2K Andor iKon-L BEX2-DD CCD camera with a pixel scale of $0.35\arcsec$, resulting in a field-of-view of $12\arcmin\times12\arcmin$ \citep{Burdanov2022}. It is a twin of the SPECULOOS-South (Section \ref{sso})  and SAINT-EX (Section \ref{saintex}) telescopes. 
The first transit was observed on UTC 2020 August 13, and the second was observed on UTC June 25 2021. Both transits were observed in the $I+z$ filter with an exposure time of 33\,s, and  we performed aperture photometry in an uncontaminated target apertures of 2.8--3.2\arcsec and a PSF {\it FWHM} of 1.4--1.6\arcsec.
Data reduction (bias, dark and flat correction) and photometric measurements were performed using the {\tt PROSE} pipeline \citep{garcia2021}.

\subsubsection{SAINT-EX} \label{saintex}
We used the SAINT-EX  telescope to observe one full transit of TOI-2084.01 on UTC July 13 2021 in the $r'$ filter with an exposure time of 141\,seconds. SAINT-EX (\href{https://www.saintex.unibe.ch/saint_ex_observatory/}{Search And characterIsatioN of Transiting EXoplanets}, \citet{Demory_AA_SAINTEX_2020})  is a 1-m F/8 Ritchey-Chretien telescope located at the Sierra de San Pedro M\'artir in Baja California, M\'exico.  SAINT-EX is equipped with a thermoelectrically cooled 2K $\times$ 2K Andor iKon-L CCD camera. 
The detector gives a field-of-view of 12\arcmin$\times$12\arcmin\ with a pixel scale of $0.35\arcsec$ per pixel. 
 We performed aperture photometry in an uncontaminated target aperture of 3.2\arcsec and a PSF {\it FWHM} of 1.4\arcsec.
Data reduction  and  photometric measurements were performed using the {\tt PROSE} pipeline \citep{garcia2021}. 

\subsubsection{TRAPPIST-North} \label{TN}

We used the 60-cm TRAPPIST-North telescope to observe one partial transit and one full transit of TOI-2084.01. TRAPPIST-North (\href{https://www.trappist.uliege.be}{TRAnsiting Planets and PlanetesImals Small Telescope}) is a 60-cm robotic telescope installed at Oukaimeden Observatory in Morocco since 2016 (\citet{Barkaoui2019_TN}, and references therein). It is equipped with a thermoelectrically cooled 2K$\times$2K Andor iKon-L BEX2-DD CCD camera with a pixel scale of 0.6\arcsec\ and a field-of-view of $20\arcmin\times20\arcmin$.
The first transit was observed on UTC January 30 2021 in the $I+z$ filter with an exposure time of 60\, seconds. 
 We took 154 science images and performed aperture photometry in an uncontaminated aperture of 7.6\arcsec and a PSF  {\it FWHM} of 3.1\arcsec.
The second transit was observed on UTC June 25 2021 in the $I+z$ filter with an exposure time of 65\, seconds. 
 We took 216 science images and performed aperture photometry in an uncontaminated aperture of 5.6\arcsec and a PSF {\it FWHM} of 3.7\arcsec. 
During that second observation of TOI-2084, the telescope underwent a meridian flip at BJD 2459391.4829. Data reduction  and  photometric measurements were performed using the {\tt PROSE} pipeline \citep{garcia2021}. 

\subsubsection{TRAPPIST-South}

Two full transits of TOI-4184.01 were observed with the TRAPPIST-South telescope. TRAPPIST-South is a 60-cm Ritchey-Chretien telescope located at ESO-La Silla Observatory in Chile, which is the twin of TRAPPIST-North (Section~\ref{TN}). It is equipped with a thermoelectrically cooled 2K$\times$2K FLI Proline CCD camera with a field of view of $22\arcmin\times22\arcmin$ and pixel-scale of 0.65\arcsec/pixel \citep{Jehin2011,Gillon2011}. The first transit was observed on UTC August 2 2021, and the second transit was observed on UTC September 25 2021. Both transits were observed in the $I+z$ filter with an exposure time of 150\,s, and  we performed aperture photometry in an uncontaminated target apertures of 3.5--6.2\arcsec and a PSF {\it FWHM} of 2.4--2.7\arcsec. During the second transit of TOI-4184.01, the telescope underwent a meridian flip at BJD = 2459478.8226. Data reduction and photometric measurements were performed using the {\tt PROSE} pipeline \citep{garcia2021}. 

\subsubsection{LCOGT-2.0m MuSCAT3}

We used the Las Cumbres Observatory Global Telescope (LCOGT; \cite{Brown_2013}) 2.0-m
Faulkes Telescope North at Haleakala Observatory in Hawaii to observe two transits of TOI-2084.01 simultaneously in Sloan-$g',r',i'$ and Pan-STARRS $z$-short filters. The first (full) transit was observed on UTC May 19 2021, and the second (partial) transit was observed on UTC May 26 2021. We used uncontaminated $4\arcsec$ target apertures to extract the stellar fluxes. The telescope is equipped with the MuSCAT3 multi-band imager \citep{Narita_2020SPIE11447E}. The raw data were calibrated by the standard LCOGT {\tt BANZAI} pipeline \citep{McCully_2018SPIE10707E}, and photometric measurements were extracted using {\tt AstroImageJ}\footnote{{\tt AstroImageJ:}~\url{https://www.astro.louisville.edu/software/astroimagej/}} \citep{Collins_2017}.

\subsubsection{Las Cumbres Observatory CTIO-1.0m and SAAO-1.0m}

We used the Las Cumbres Observatory Global Telescope (LCOGT; \cite{Brown_2013}) 1.0-m network to observe four full transits of TOI-4184.01 in the Sloan-$i'$ and $g'$ filters. The telescopes are equipped with 4096x4096 SINISTRO Cameras, having an image scale of $0.389\arcsec$ per pixel and a Field-Of-View of 26'x26'. The raw data were calibrated by the standard LCOGT {\tt BANZAI} pipeline \citep{McCully_2018SPIE10707E}, and photometric measurements were extracted using {\tt AstroImageJ} \citep{Collins_2017}. Two transits were observed at Cerro Tololo Interamerican Observatory (CTIO) in Sloan-$i'$ on UTC July 28 2021 and September 20 2021, using uncontaminated $3.1\arcsec$ and $4.3\arcsec$ target apertures. Two others were observed simultaneously in the Sloan-$g'$ and $i'$ at South Africa Astronomical Observatory (SAAO) on UTC October 10 2021, using uncontaminated $3.9\arcsec$ target apertures.

\subsubsection{Danish-1.54m}

Three transits of TOI-4148.01 were observed on  UTC September 15, 20 and 25 2021 by the MiNDSTEp consortium \citep{Dominik_2010AN} using the Danish 1.54 m telescope at ESO's La Silla observatory in Chile. The instrument used was the DFOSC imager, operated with a Bessell~$I$  filter for two transits and a Bessell~$R$  filter for the third. In this setup, the CCD covers a  field of view of $13.7'{\times}13.7'$ with a pixel scale of $0.39''$ pixel$^{-1}$. The images were unbinned and windowed for the  first transit, resulting in a dead time between consecutive images of 10\,s; however, in an effort to improve the SNR of the target PSF, the remaining transits used $2\times2$ binning and no windowing (to obtain a greater selection of comparison stars), resulting in a dead time between consecutive images of 13\,s. The exposure times were 60\,s for all images and transits. Due to the target being quite faint ($V= 17^{th}$mag, $I=14^{th}$mag) and with the presence of close nearby sources (both point and extended) the telescope was marginally defocused and autoguiding was maintained through all observations. The amount of defocus applied caused the resulting PSFs to have a diameter of $\approx10$ pixels for all nights.

We reduced the Danish 1.54-m telescope data using the {\sc DEFOT} pipeline \citep{Southworth_2009MNRAS,Southworth_2014MNRAS}. Aperture photometry was performed with an IDL implementation of {\sc DAOPHOT} \citep{Stetson_1987PASP}, with the addition of image motion tracking by cross-correlation with a reference image to produce a differential magnitude light curve. The light curve was produced after simultaneously  fitting a  first-order polynomial to the out of transit data. The aperture sizes and number of suitable comparison stars were adjusted to obtain the lowest baseline scatter; this method affects the scatter in the transit data but does not significantly impact the light curve shape. The timestamps from the fits files were converted to the BJD$_{\rm TDB}$ time-scale using routines from \cite{Eastman_2010PASP}.

\subsubsection{ExTrA}

The ExTrA facility \citep{Bonfils_2015SPIE_ExTrA}, located at La Silla observatory, consists of a near-infrared (0.85--1.55 $\mu m$; NIR) multi-object spectrograph fed by three 60-cm telescopes. Five fiber positioners at the focal plane of each telescope pick up light from the target and four comparison stars. We observed one full transit of TOI-4184\,b on UTC September 15 2021  with two telescopes using the $8''$ aperture fibers. We used the spectrograph's low resolution mode (R $\sim$20) and 60-second exposures. We also observed 2MASS J02542961-7941578, 2MASS J03025970-7941390, 2MASS J03025068-7918174, and 2MASS J02581731-7913567, with J-magnitudes \citep{Skrutskie_2006AJ_2MASS} and effective temperatures \citep{Gaia_Colaboratio_2018AA} similar to TOI-4184, for use as comparison stars. The resulting ExTrA data were analyzed using custom data reduction software.

\begin{table*}[!h]
 \begin{center}
 {\renewcommand{\arraystretch}{1.4}
 \begin{tabular}{l c c c c c ccc}
 \toprule
Planet & Date (UT) & Filter & Telescope & Exptime  & Aperture size   & FWHM & Coverage \\ 
       &           &        &           & [second]  & [arcsec]      & [arcsec] \\
 \hline
 TOI-2084.01 & Aug 13 2020 & $I+z$ & Artemis-1.0m & 33 & 2.8 & 1.4 & Egress \\
 TOI-2084.01 &  Jan 30 2021 & $I+z$ & TRAPPIST-N-0.6m & 65 & 7.6 & 3.1 & Egress  \\
 TOI-2084.01 &  May 20 2021 & Sloan-$g',r',i',zs$ & MuSCAT3-2.0m & - & 4.0 & 2.0 & Full\\
 TOI-2084.01 &  May 26 2021 & Sloan-$g',r',i',zs$ & MuSCAT3-2.0m & - & 4.0 & 2.0 &  Egress\\
 TOI-2084.01 &  Jun 25 2021 & $I+z$ & Artemis-1.0m & 33 & 3.2 & 1.6 & Full\\
 TOI-2084.01 &  Jun 25 2021 & $I+z$ & TRAPPIST-N-0.6m & 65 & 5.6 & 3.7 & Full \\
 TOI-2084.01 &  Jul 13 2021 & Sloan-$r'$ & SAINT-EX-1.0m & 141 & 3.2 & 1.4 & Egress\\
 \hline
 TOI-4184.01 &  Jul 28 2021 & ip & LCO-CTIO-1.0m & 240 & 3.1 & 1.3 & Full \\
 TOI-4184.01 &  Aug 02 2021 & $I+z$ & TRAPPIST-S-0.6m & 150 & 3.5 & 2.4 & Full \\
 TOI-4184.01 &  Sep 15 2021 & $I_C$ & Danish-1.54m & 60 & 2.8 & 1.3 & Full \\
 TOI-4184.01 &  Sep 15 2021 & $1.21 \mu m$ & ExTrA-0.6m & 60 & 4.0 & 1.5 & Full \\
 TOI-4184.01 &  Sep 20 2021 & $I_C$ & Danish-1.54m & 60 & 5.5 & 4.5 & Full \\
 TOI-4184.01 &  Sep 20 2021 & $I+z$ & TRAPPIST-S-0.6m & 150 & 6.2 & 2.7 & Full + Flip \\
 TOI-4184.01 &  Sep 20 2021 & Sloan-$i'$ & LCO-CTIO-1.0m & 240 & 4.3 & 2.2 & Full \\
 TOI-4184.01 &  Sep 25 2021 & Sloan-$z'$ & SPECULOOS-S-1.0m & 42 & 3.9 & 1.7 & Full\\
 TOI-4184.01 &  Sep 25 2021 & Rc & Danish-1.54m & 100 & 4.0 & 2.8 & Full\\
 TOI-4184.01 &  Oct 10 2021 & Sloan-$g',i'$ & LCO-SAAO-1.0m & 400,240 & 3.9 & 2.3 & Full\\
 \hline
 \end{tabular}}
 \caption{Table shows the observational parameters:  date of observation, filter used, telescope, exposure time(s), photometric aperture size, and FWHM of the point-spread function.}
 \label{obs_table}
 \end{center}
\end{table*}

\begin{figure*}[!h]
	\centering
	\includegraphics[scale=0.5]{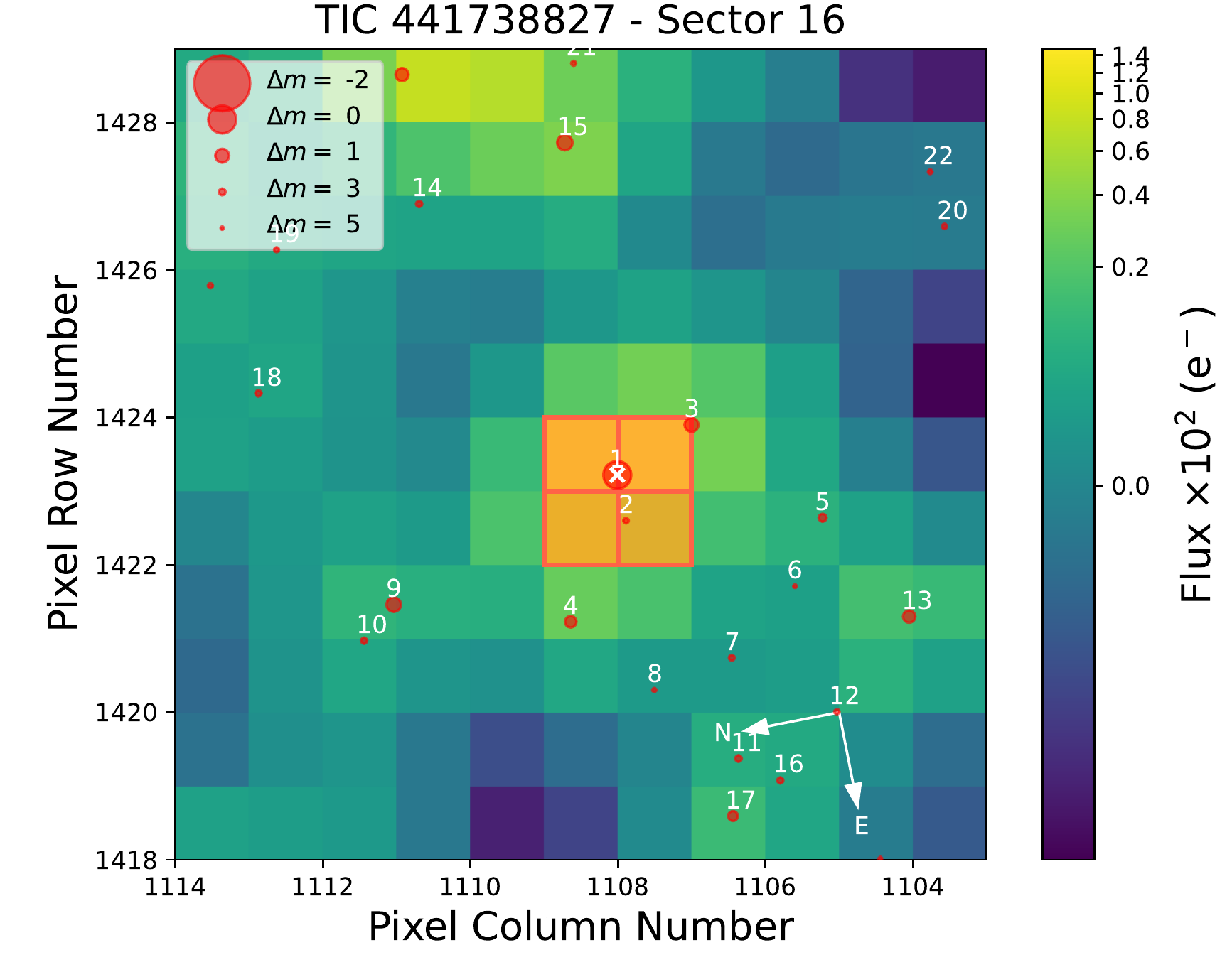}
	\includegraphics[scale=0.5]{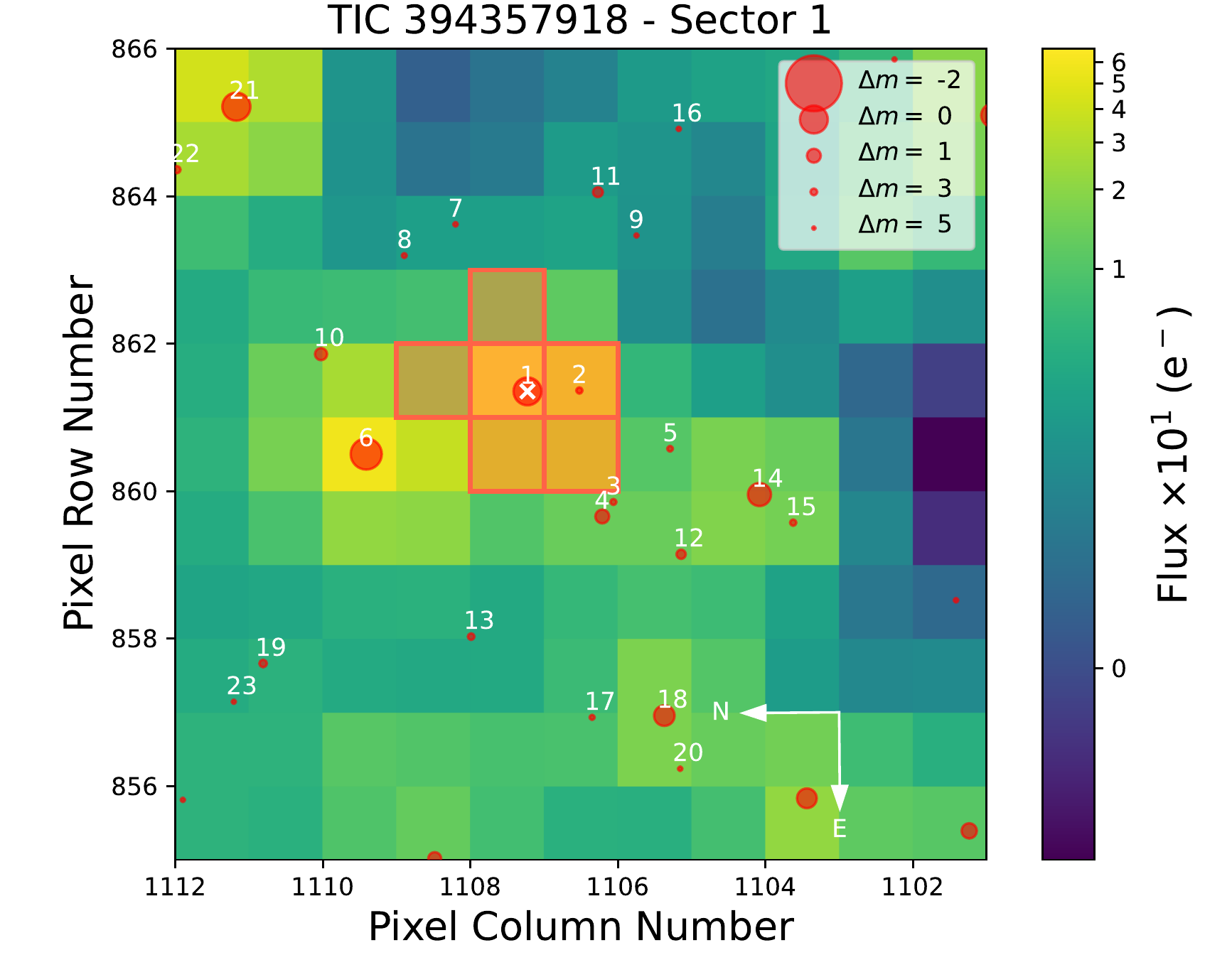}
	\caption{\emph{TESS} target pixel file images of TOI-2084 observed in Sector 16 (left panel) and TOI-4184 observed in Sector 1 (right panel), made by \href{https://github.com/jlillo/tpfplotter}{\tt tpfplotter} \citep{Aller_2020A&A}. Red dots show the location of Gaia DR3 sources, and the red shaded region shows the photometric apertures used to extract the photometric measurements.} 
	\label{Target_pixel}
\end{figure*}

\begin{figure*}
	\centering
	\includegraphics[scale=0.6]{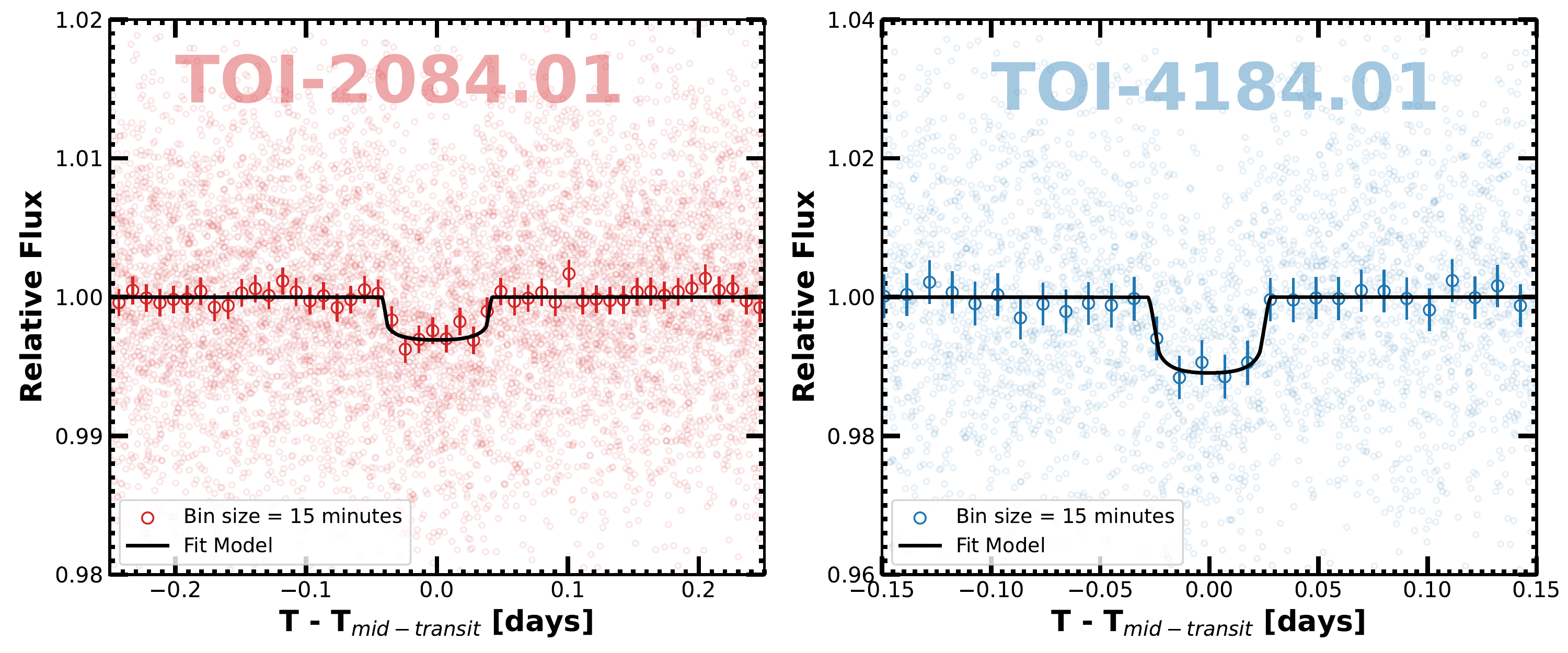}
	\caption{
	Phase-folded \emph{TESS} transit light curves of TOI-2081.01 (left) and TOI-4184.01 (right). The points with the error bar are data binned to 15 minutes.
	}
	\label{plot_lightcurves_TESS}
\end{figure*}

\begin{figure*}
	\includegraphics[scale=0.45]{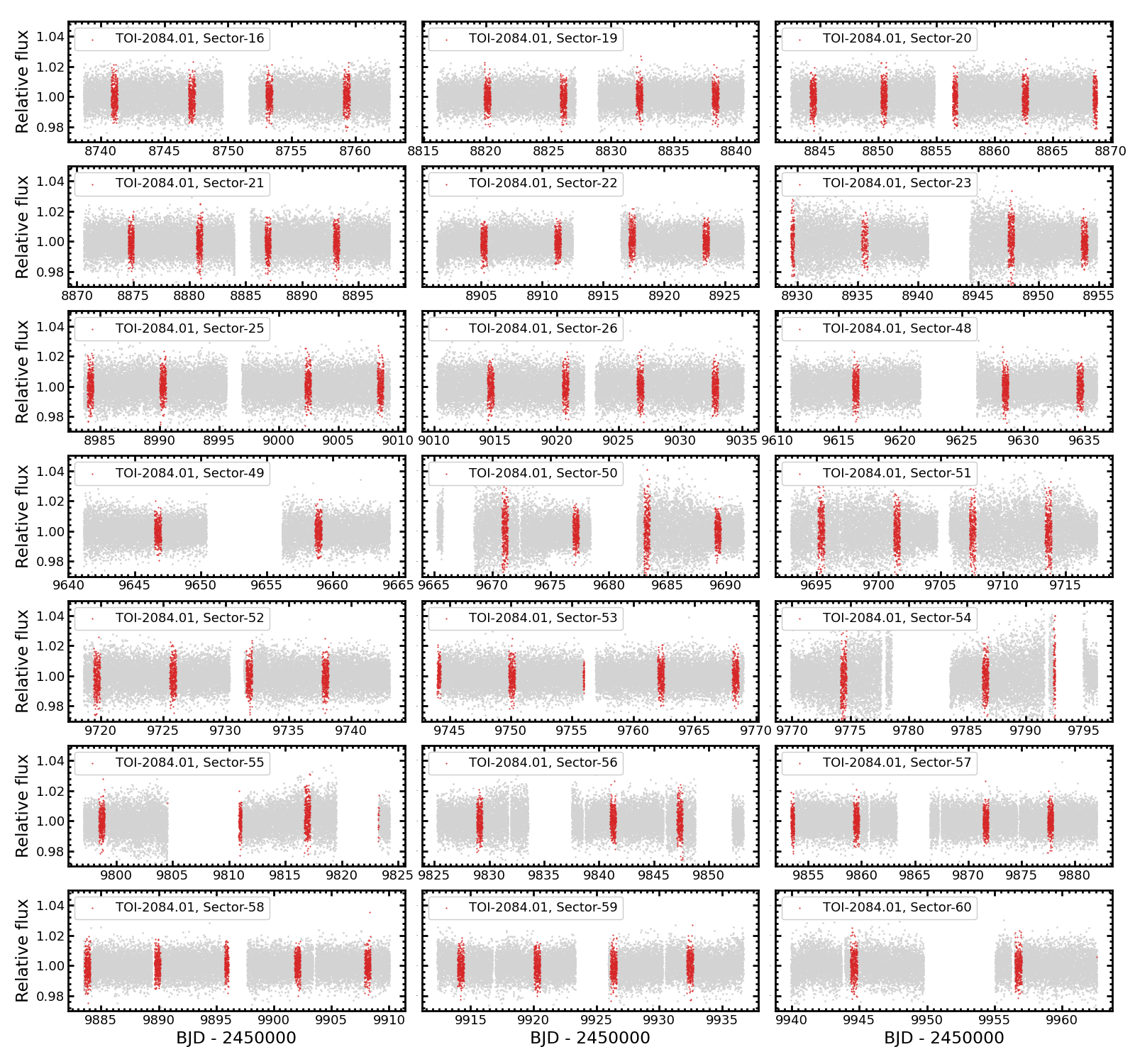}
	\includegraphics[scale=0.447]{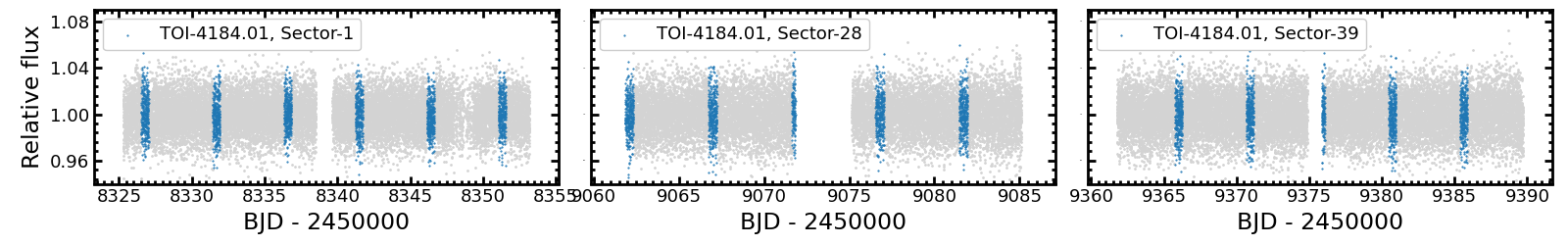}
	\caption{\emph{TESS} photometric data of TOI-2084.01 and TOI-4184.01. The gray points show the PDSAP fluxes obtained from the SPOC pipeline. The red and blue points correspond to the location of the transit for the candidates TOI-2084.01 and TOI-4184.01, respectively.}
	\label{TOI_2084_4184_TESS_LC}
\end{figure*}

\begin{figure*}
	\centering
	\includegraphics[scale=0.7]{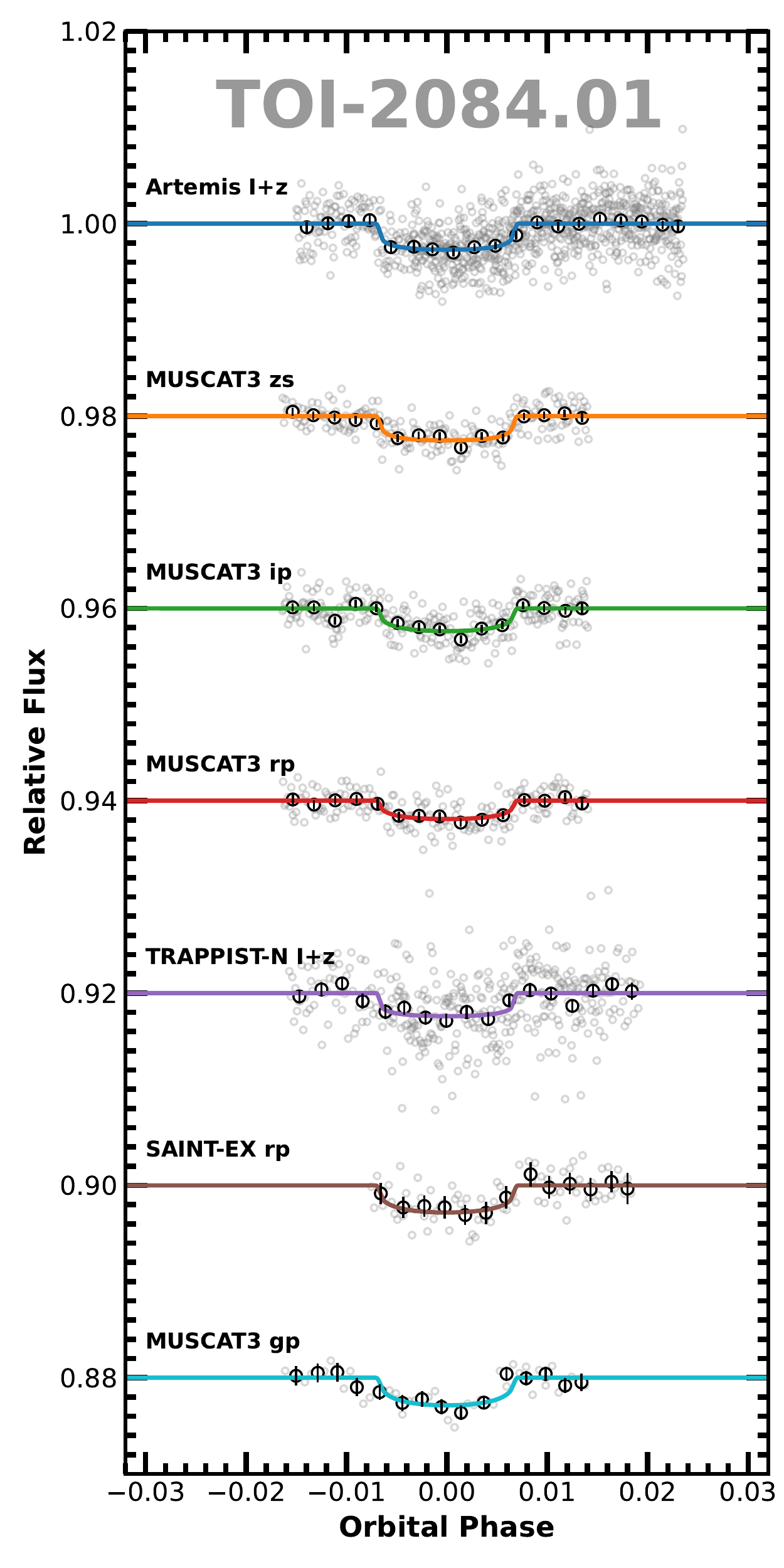}
	\includegraphics[scale=0.7]{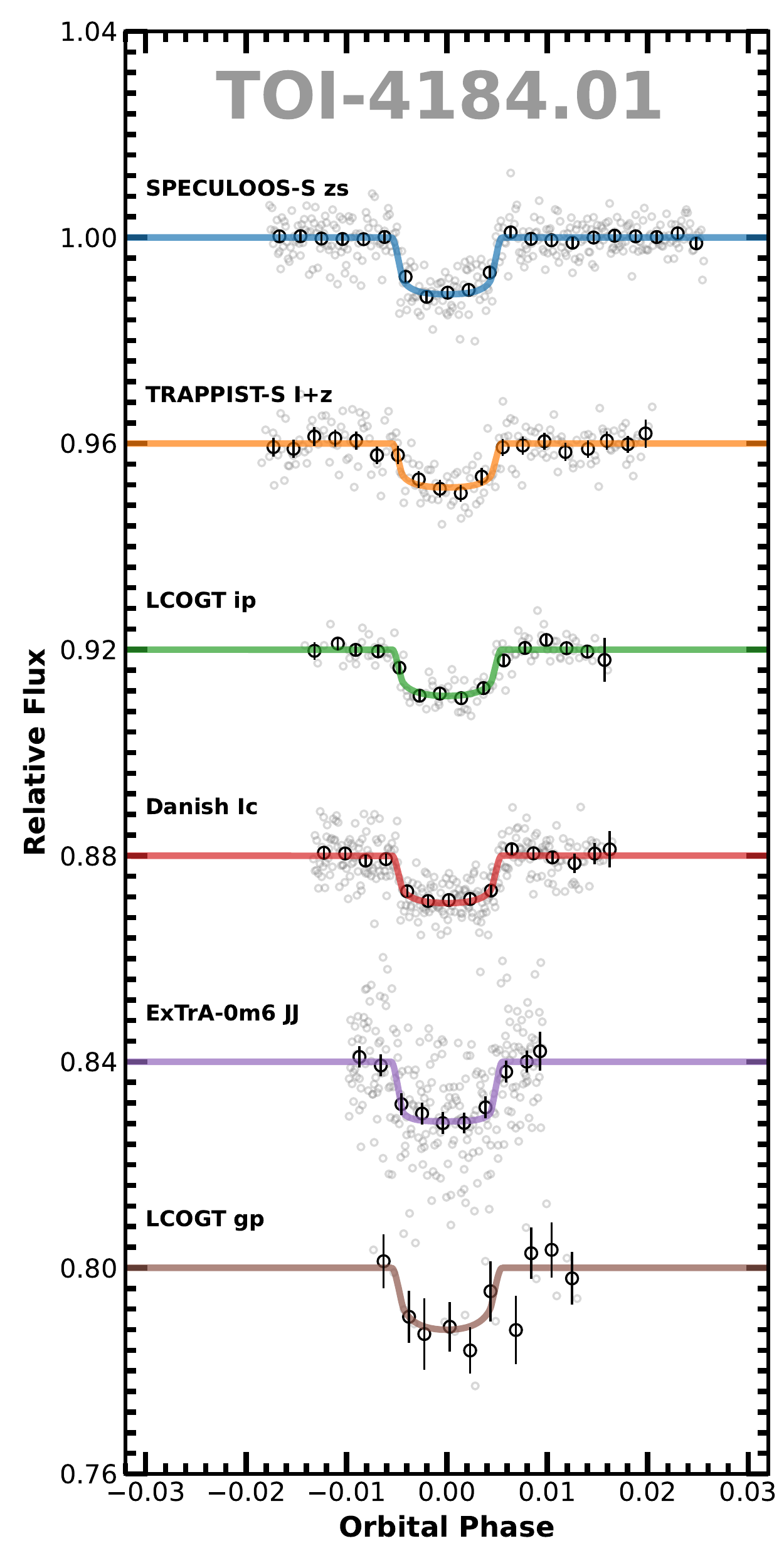}
	\caption{
	 Ground-based photometric light curves of TOI-2081.01 (left) and TOI-4184.01 (right). The gray points are unbinned data and the black points are data binned to 10 minutes. The coloured lines are the best-fitting transit model. The light curves are shifted along the y-axis for visibility.}
	\label{plot_lightcurves_GB}
\end{figure*}

\begin{figure*}
	\centering
	\includegraphics[scale=0.6]{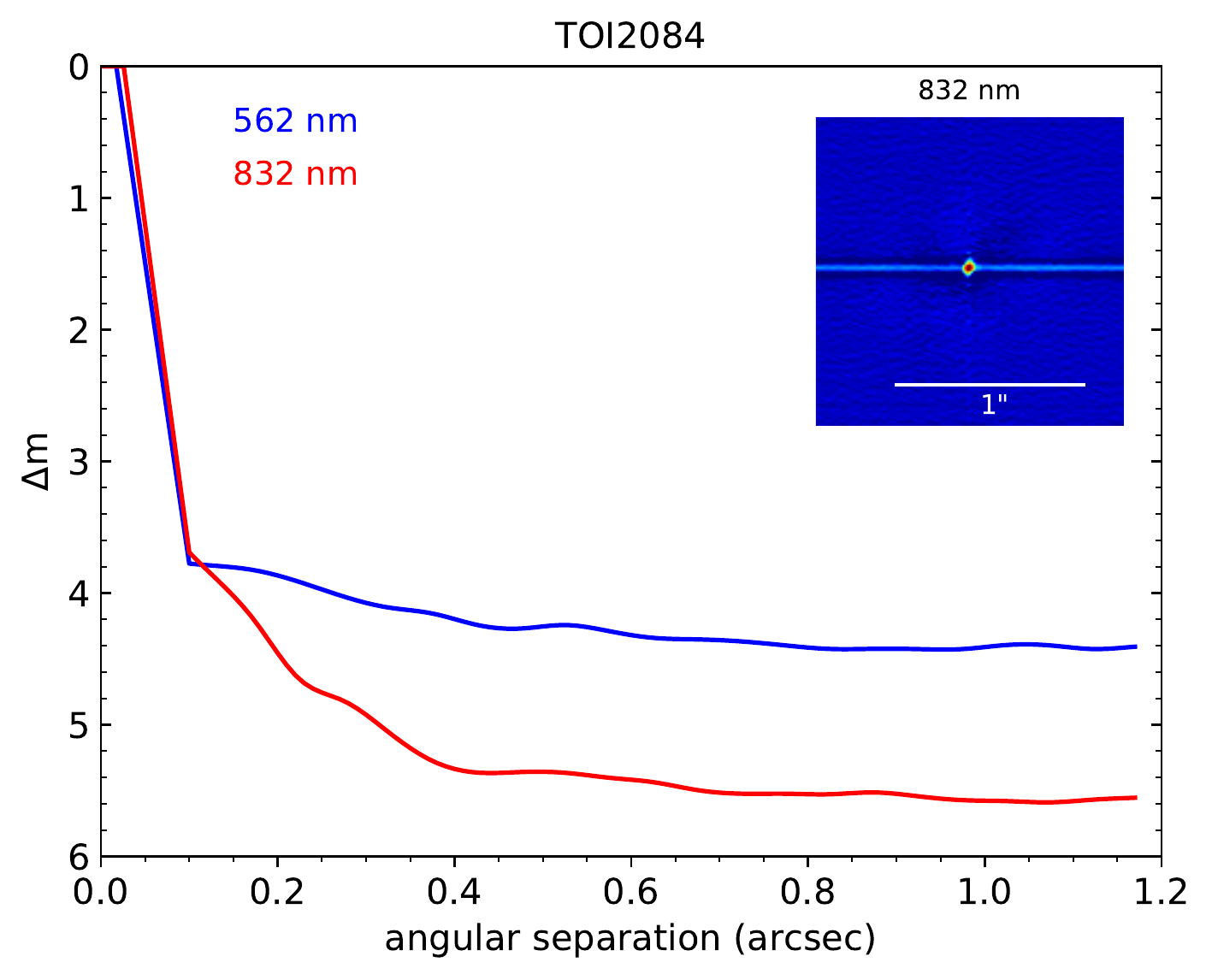}
	\includegraphics[scale=0.6]{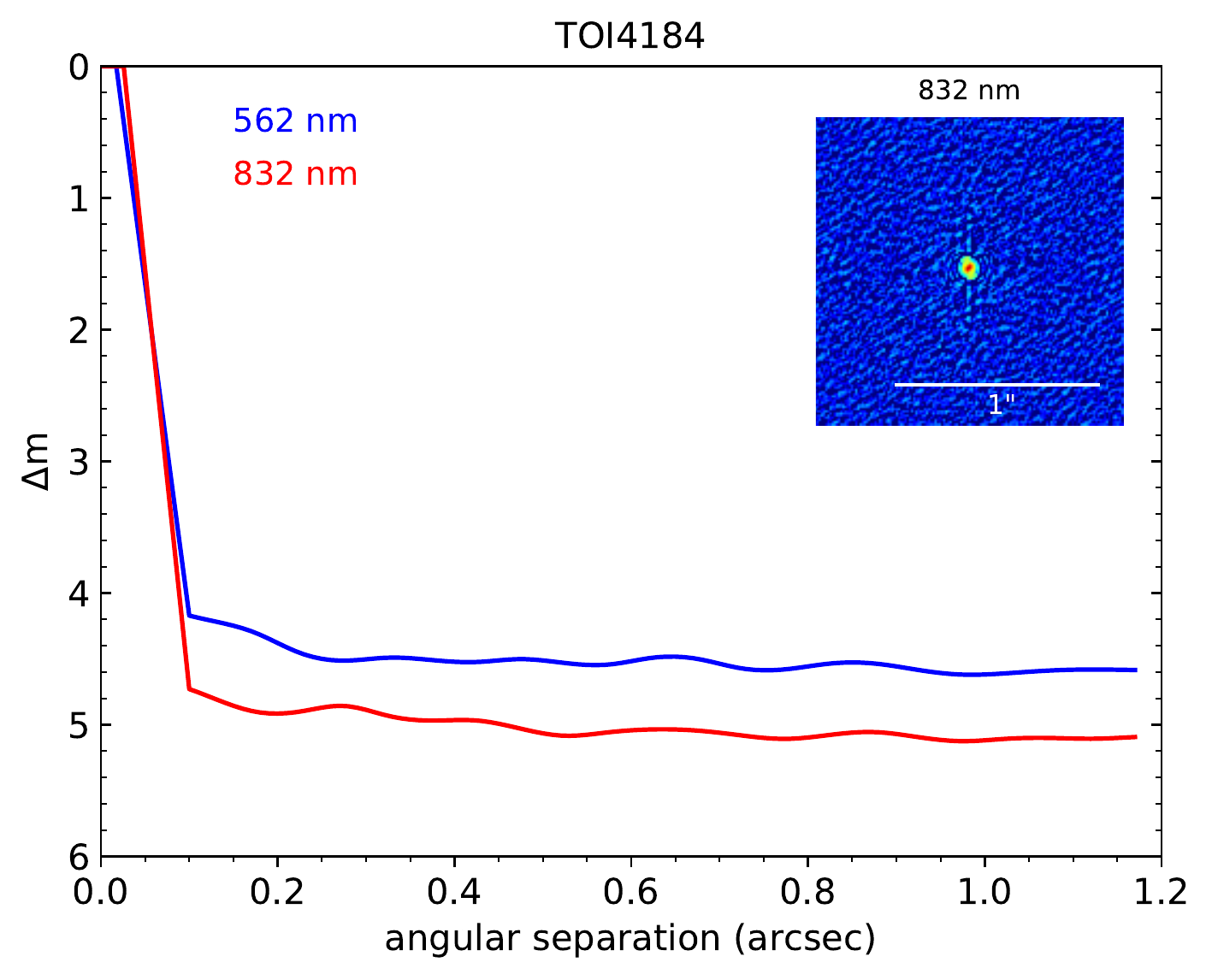}
	\caption{{\it Left panel:} Gemini-North/Alopeke high resolution image of TOI-2084 observed on UTC June 24 2021. {\it Right panel:} Gemini-South/Zorro high-resolution image of TOI-4184 observed on UTC December 23 2021. TOI-2084 and TOI-4184 are both single stars with no companion brighter than 4-6 magnitudes below that of the target star.}
	\label{toi_2084_4184_Gemini}
\end{figure*}

\begin{figure*}
	\centering
	\includegraphics[scale=0.5]{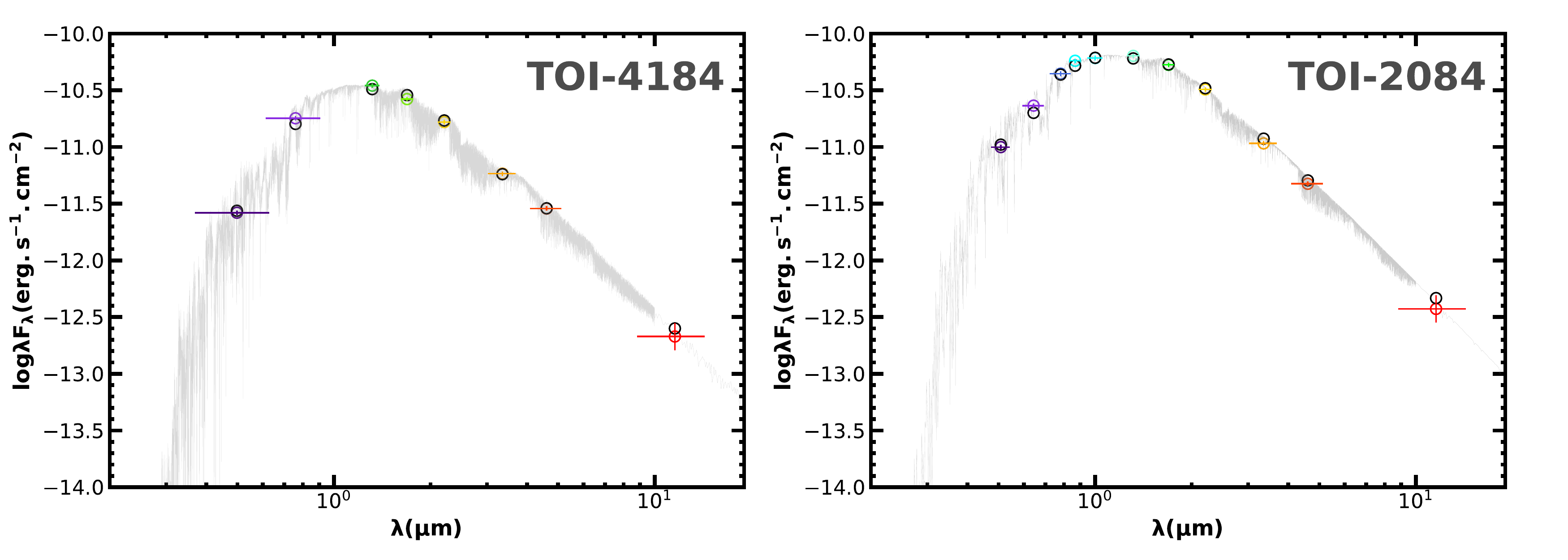}
	\caption{Spectral Energy Distribution (SED) fit of TOI-2084 (right) and TOI-4184 (left). The gray curves are the best-fitting NextGen atmosphere model, coloured symbols with error-bars are the observed fluxes, and black symbols are the model fluxes.}
	\label{SED_plots}
\end{figure*}

\begin{figure*}
	\centering
	\includegraphics[scale=0.5]{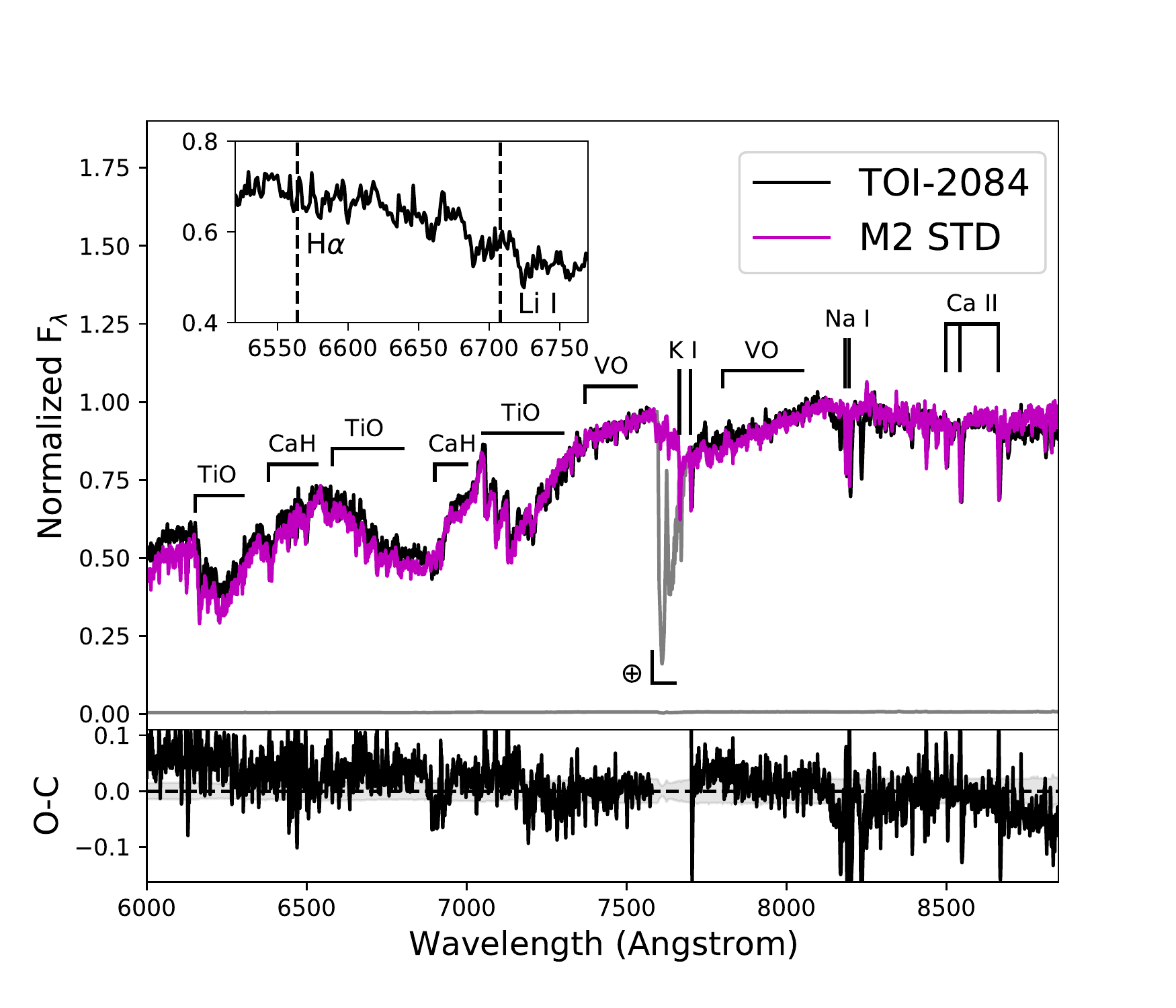}
	\includegraphics[scale=0.5]{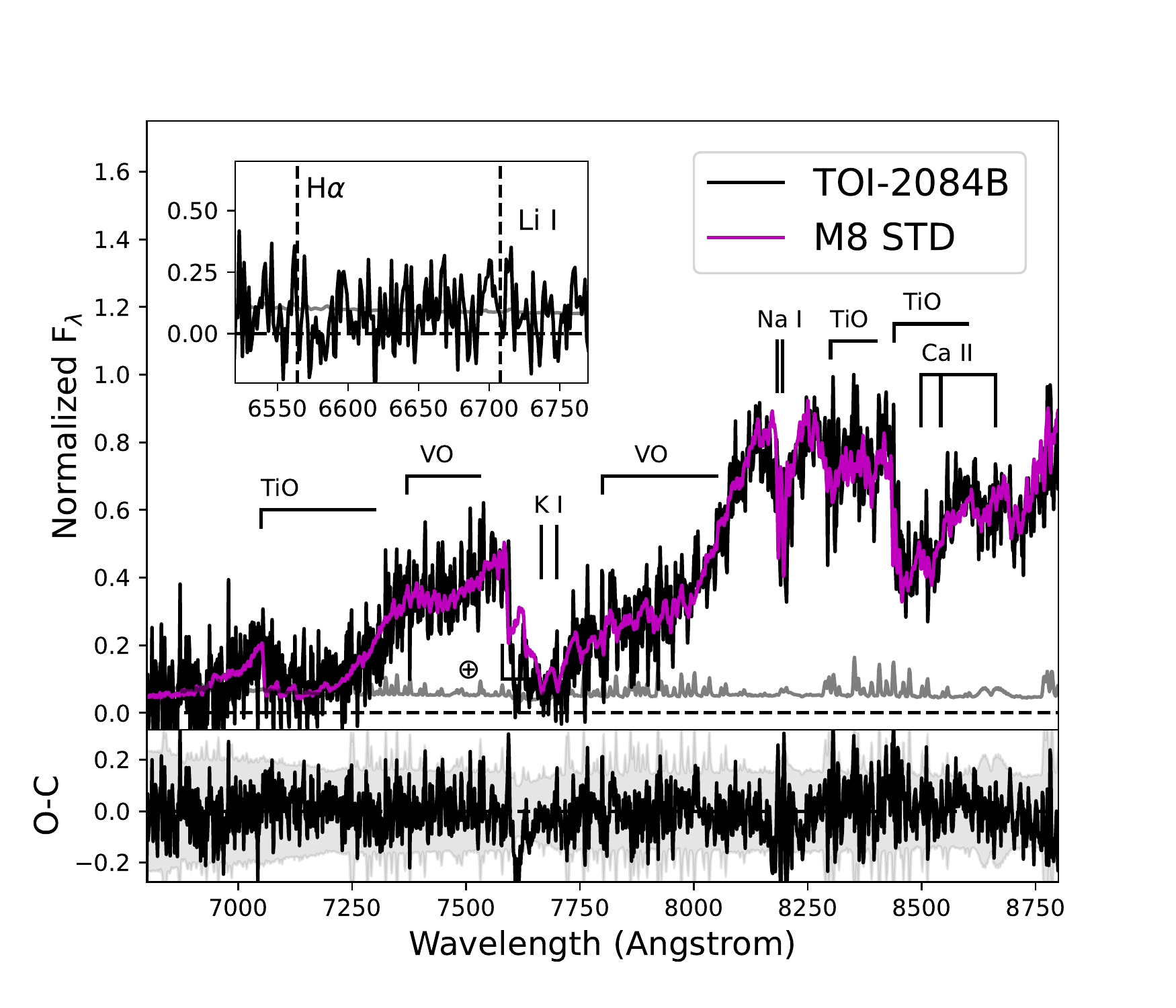}
	\caption{
	    Shane/Kast red optical spectra (black lines) of TOI-2084 (left) and its wide stellar companion TOI-2084B (right) compared to best-fit M2 and M8 SDSS spectral templates from \citet[magenta lines]{2007AJ....133..531B}. The lower panels display the difference between these spectra (black line) compared to the $\pm$1$\sigma$ measurement uncertainty (grey band). 
     Key features are labeled, including the strong telluric O$_2$ band at 7600~{\AA} ($\oplus$). 
     Inset boxes show close-ups of the region around the 6563~{\AA} H$\alpha$ and 6708~{\AA} Li~I lines.
	    }
	\label{fig:stellar_spectra_OPT}
\end{figure*}

\begin{figure}
	\centering
	\includegraphics[scale=0.56]{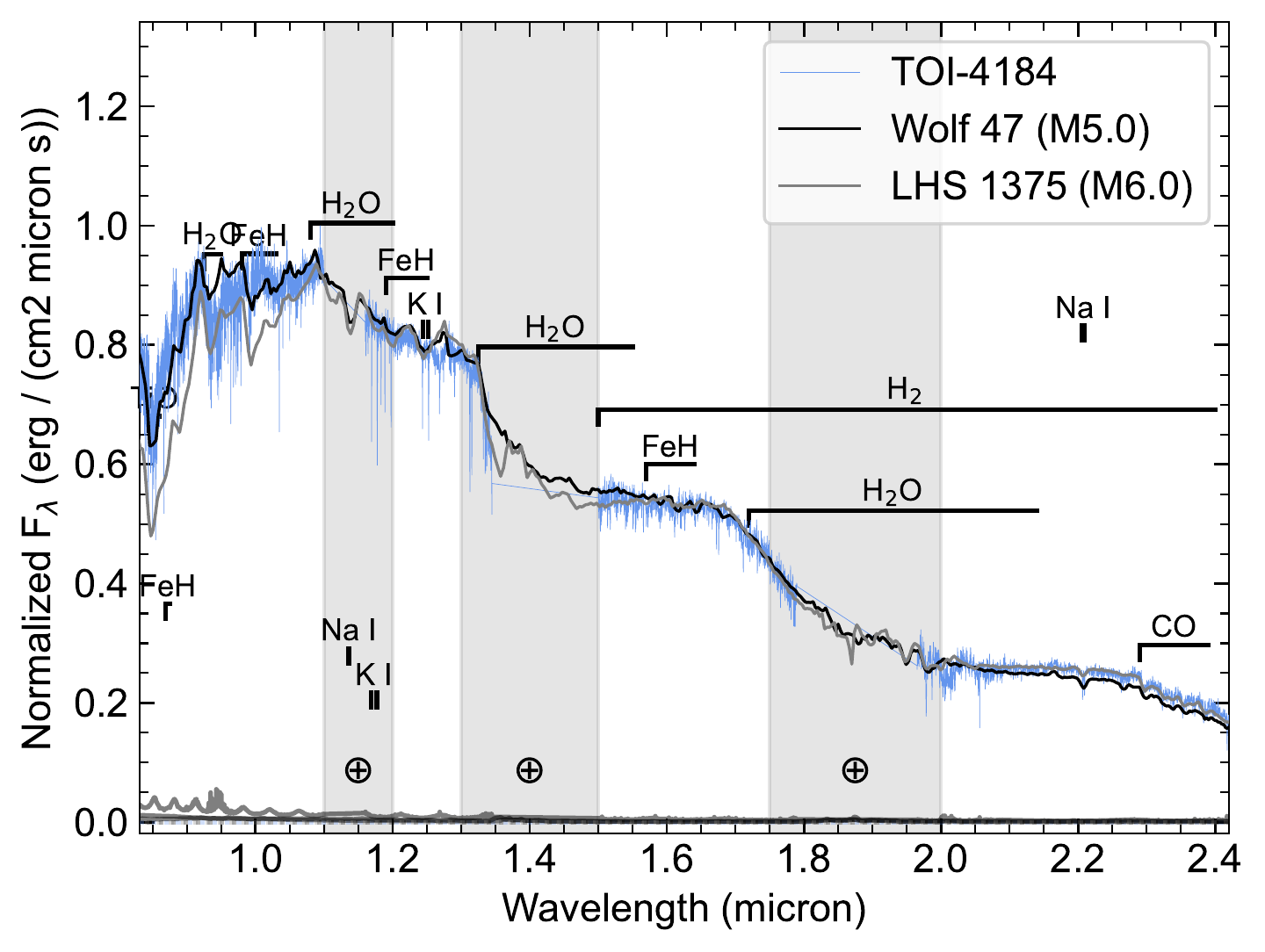}
	\caption{
     Magellan/FIRE spectrum of TOI-4184.
	    The SpeX Prism spectrum of the M5.0 standard Wolf 47 and M6.0 standard LHS 1375 \citep{Kirkpatrick2010} are shown for comparison.
	    Strong M dwarf spectral features are indicated, and high-telluric regions are shaded.
	    Lines along the bottom of the plot give the uncertainties associated with the spectra.
	}
	\label{fig:stellar_spectra_NIR}
\end{figure}

\begin{figure}
	\includegraphics[scale=0.34]{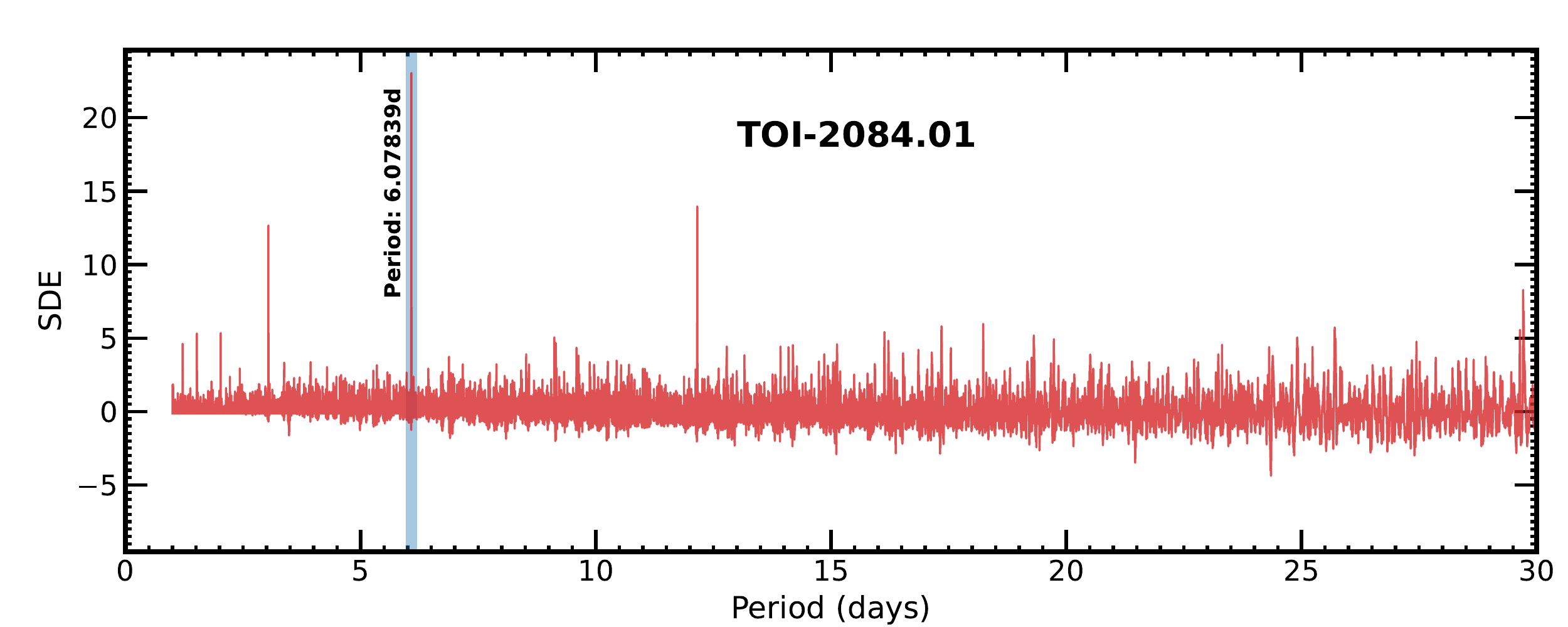}
	\includegraphics[scale=0.34]{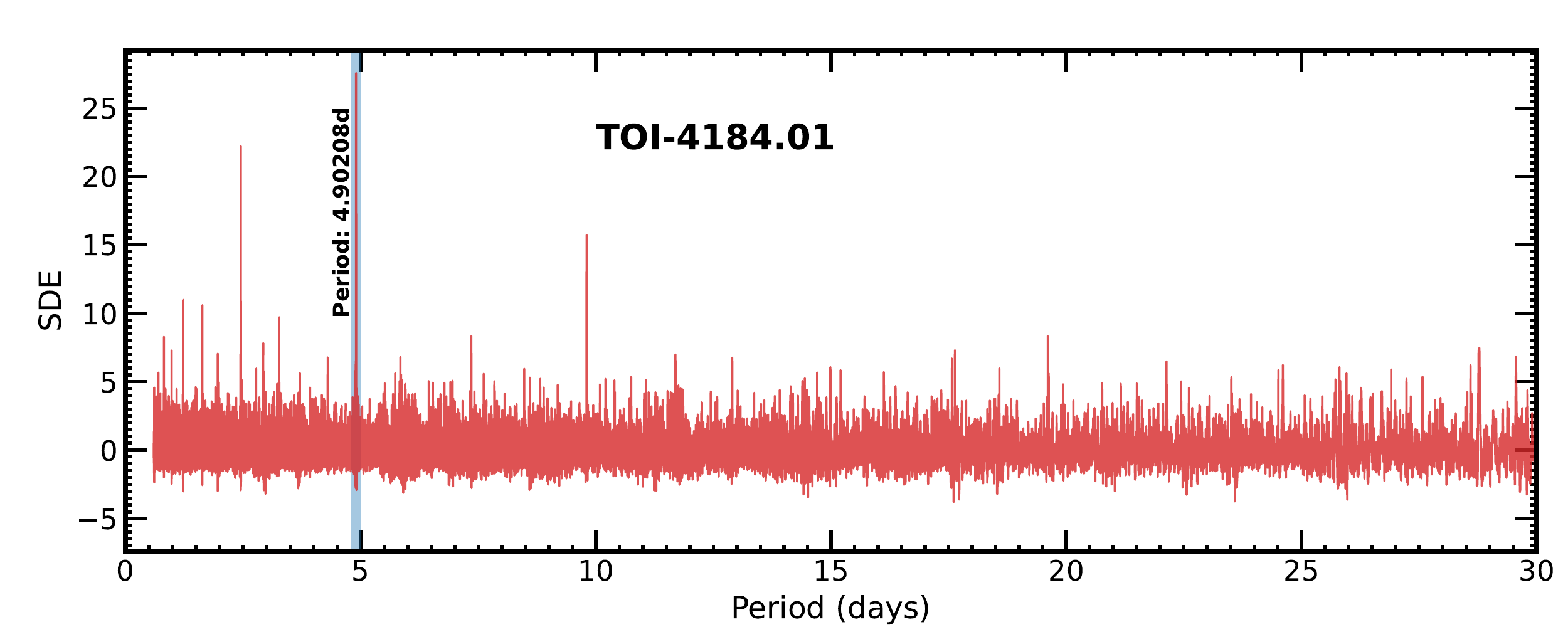}
	\caption{Generalised Lomb-Scargle periodogram (GLS, \cite{Zechmeister_2009A&A}) for TOI-2084.01 (top) and TOI-4184.01 (bottom) \emph{TESS} data.}
	\label{plot_periodogram}
\end{figure}

\begin{figure}
	\centering
	\includegraphics[scale=0.25]{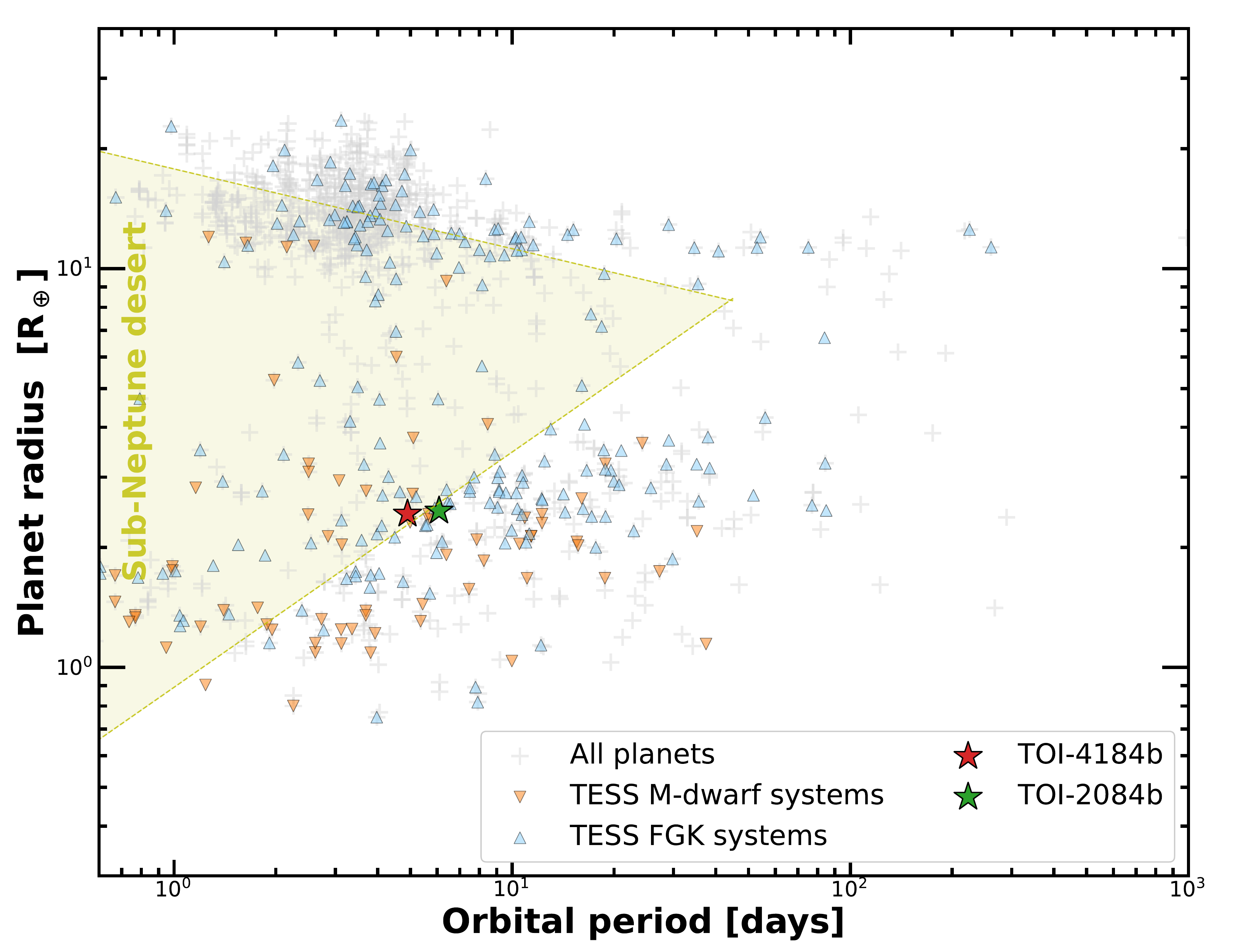}
	\caption{Period--Radius diagram of known transiting exoplanets from \href{https://exoplanetarchive.ipac.caltech.edu/}{\it NASA Archive of Exoplanets}. The blue and orange data points correspond to the \emph{TESS} FGK and M dwarf systems, respectively.
	The green and red stars show TOI-2084\,b TOI-4184\,b, respectively. The yellow region shows the boundaries of the sub-Neptune-desert determined by \cite{Mazeh_2016AandA_Jovian_desert}.}
	\label{Desert_Neptune}
\end{figure}

\subsection{Spectroscopy}

\subsubsection{Shane/Kast Optical Spectroscopy}


We obtained an optical spectrum of TOI-2084  and its co-moving companion (see below) on UTC November 13 2021 using the Kast double spectrograph \citep{kastspectrograph} mounted on the 3-m Shane Telescope at Lick Observatory in clear conditions. Six exposures of 600\,s each was obtained  of both sources TOI-2084  simultaneously using the 600/7500 grism and 1$\farcs$5-wide slit, providing 6000--9000~{\AA} wavelength coverage at an average resolution of $\lambda/\Delta\lambda$ = 1900. We also observed the flux calibrator Feige 110 later that night \citep{1992PASP..104..533H,1994PASP..106..566H}. Data were reduced using the kastredux package\footnote{kastredux:~\url{https://github.com/aburgasser/kastredux}.}, which included image reduction, boxcar extraction of the one-dimensional spectra, wavelength calibration, and flux calibration. No correction for telluric absorption was applied. The final  spectra have median signals-to-noise of 125 (TOI-2084) and 12 (TOI-2084B) around 8400~{\AA}, with a wavelength accuracy of 0.26~{\AA} (12~km/s).

\subsubsection{Magellan/FIRE Spectroscopy}
We obtained a spectrum of TOI-4184 with the FIRE spectrograph \citep{Simcoe2008} on the 6.5-m Magellan Baade Telescope on  UTC September 23, 2021.
We used the high-resolution echellette mode with the $0\farcs60$ slit, providing a 0.82--2.51\,{\micron} spectrum with a resolving power of $R{\sim}6000$.
We collected a single ABBA nod sequence (4 exposures) with integration times of 95.1\,s per exposure, giving a total exposure time of 380.4\,s.
After the science exposures, we collected a pair of 15-s exposures of the A0\,V star HD\,45039 for flux and telluric calibrations followed by a pair of 10-s arc lamp exposures and a set of 10 1-s flat-field exposures.
We reduced the data using the \texttt{FIREHOSE} pipeline\footnote{FIREHOSE:~\url{https://github.com/rasimcoe/FIREHOSE}}.
The final spectrum (\autoref{fig:stellar_spectra_NIR}) has a median SNR of 77, with peaks in the J, H, and K bands of 120--140.

\subsection{High-Resolution Imaging from Gemini-8m0} \label{High_Reso_Imag}
TOI-2084 was observed on  UTC June 24 2021 using the 'Alopeke speckle instrument on the Gemini North 8-m telescope and TOI-4184 was observed on UTC December 23 2021 using the Zorro speckle instrument on the Gemini South 8-m telescope (see \cite{Scott_2021FrASS}).
'Alopeke and Zorro provide simultaneous speckle imaging in two bands (562 nm and 832 nm) with output data products including a reconstructed image with robust contrast limits on companion detections \citep[e.g.,][]{Howell_2016ApJ}. A total of 13/11 sets of $1000\times0.06$ sec exposures were collected for TOI-2084/TOI-4184 and subjected to Fourier analysis in our standard reduction pipeline \citep[see][]{Howell_2011AJ}. \autoref{toi_2084_4184_Gemini} shows our final 5$\sigma$ contrast curves and the 832\,nm reconstructed speckle images. We find that TOI-2084 and TOI-4184 are both single stars with no companion brighter than about 4--6 magnitudes below that of the target star from the diffraction limit (20 mas) out to 1.2". At the distance of TOI-2084/TOI-4184 (d=114/69 pc), these angular limits correspond to spatial limits of 2.3 to 137 au (TOI-2084) and 1.4 to 83 au (TOI-4184).

\begin{table*}
\centering
	{\renewcommand{\arraystretch}{1.4}
		\begin{tabular}{lccc}
			\hline
			\hline
			\multicolumn{3}{c}{  Star information}   \\
			\hline
			\hline
			Parameter & TOI-2084 &  TOI-4184 & Source   \\
			\hline
			{\it Identifying information:} & & \\
			TIC            & 441738827 & 394357918 &  \\
			GAIA DR2 ID  & 1652137995942479744 & 4620574887039870720 & \\
			2MASS ID      & 2MASS J17170094+7244486 &  2MASS J02551841-7924554 & \\
			\hline
			{\it Parallax  and distance:} &   \\
			RA [J2000]     &  17:17:01.09  & 02:55:18.83 &  (1) \\
			Dec [J2000]    & +72:44:49.28  & -79:24:52.93 &  (1)\\
			Plx [$mas$] & $8.750 \pm 0.017$ & $14.451 \pm 0.027$ &  (1)\\
                $\mu_{RA}$ [mas yr$^{-1}$] & $47.73 \pm 0.02$ & $79.23 \pm 0.03$ & (1) \\
                $\mu_{Dec}$ [mas yr$^{-1}$] & $36.76 \pm 0.02$ & $165.93 \pm 0.03$ & (1) \\
			Distance [pc]  & $114.29 \pm 0.22$ & $69.20 \pm 0.13$  & (1)\\
			\hline
			{\it Photometric properties:} & \\
			TESS$_{\rm mag}$           &  $13.326 \pm 0.007$  & $14.261 \pm 0.007$ & (2)  \\
			$V_{\rm mag}$ [UCAC4]       & $15.115 \pm 0.065$  & $17.12 \pm 0.20$  & (3) \\
			$B_{\rm mag}$ [UCAC4]       & $16.668 \pm 0.033$  & -  &  (3) \\
			$J_{\rm mag}$ [2MASS]       & $11.961 \pm 0.021$  &  $12.617 \pm 0.023$ & (4) \\
			$H_{\rm mag}$ [2MASS]       & $11.356 \pm 0.018$  &  $12.111 \pm 0.026$ &  (4) \\
			$K_{\rm mag}$ [2MASS]       & $11.148 \pm 0.020$  &  $11.867 \pm 0.025$ &  (4) \\			
			$G_{\rm mag}$ [Gaia DR3]   & $14.4096 \pm 0.0005$  &  $15.5939 \pm 0.0008$ & (1)  \\
			$W1_{\rm mag}$ [WISE]       & $11.017 \pm 0.023$ &  $11.685 \pm 0.023$ & (5) \\
			$W2_{\rm mag}$ [WISE]       & $10.927 \pm 0.020$  &  $11.472 \pm 0.020$ & (5) \\
			$W3_{\rm mag}$ [WISE]       & $10.763 \pm 0.061$  & $11.371 \pm 0.120$  &  (5)\\
			\hline
			\multicolumn{2}{l}{\it Spectroscopic and derived parameters}  \\
			$T_{\rm eff}$ [K]              & $\bf 3550 \pm 50$  & $\bf 3225 \pm 75$ &  this work\\
			$\log g_\star$ [dex]           & $\bf 4.75 \pm 0.05$  & $\bf 5.01 \pm 0.04$ & this work\\
			$[Fe/H]$ [dex]                 & $\bf -0.13\pm0.20$  & $\bf -0.27 \pm 0.09$ & this work\\
			$M_\star$  [$M_\odot$]         & $\bf 0.49 \pm 0.03$  & $\bf 0.240 \pm 0.012$ & this work\\
			$R_\star$  [$R_\odot$]         & $\bf 0.475 \pm 0.016$  & $\bf 0.242 \pm 0.013$ & this work\\
			$L_{\rm bol}$  [erg s$^{-1}$ cm$^{-2}$]         & $ 7.90 \pm 0.28\times 10^{-11}$  &  $3.81 \pm 0.18\times 10^{-11}$ & this work\\
			$Av$ [mag]    & $0.02 \pm 0.02$  & $0.094 \pm 0.07$   & this work\\
			$L_\star$  [$L_\odot$]         & $0.0322_{-0.0025}^{+0.0028}$  & $0.00544_{-0.00073}^{+0.00082}$ & this work\\
			$\rho_\star$  [$\rho_\odot$]   & $4.21_{-0.16}^{+0.17}$  &  $16.32_{-1.09}^{+1.25}$ & this work\\
			$Age$  [Gyr]                   & $7.5 ^{+4.4}_{-5.1}$  & $6.7 ^{+4.9}_{-4.7}$ & this work\\
			Spectral type                  &   M2$\pm$0.5 (optical) & M5.5 $\pm$ 0.5 (NIR) & this work\\
			\hline
	\end{tabular} }
	\caption{Astrometry, photometry, and spectroscopy stellar properties of TOI-2084 and TOI-4184. 
	{\bf (1):} Gaia EDR3 \cite{Gaia_Collaboration_2021A&A}; 
	{\bf (2)} \emph{TESS} Input Catalog \cite{Stassun_2018AJ_TESS_Catalog}; 
	{\bf (3)} UCAC4 \cite{Zacharias_2012yCat.1322};
	{\bf (4)} 2MASS \cite{Skrutskie_2006AJ_2MASS};
	{\bf (5)} WISE \cite{Cutri_2014yCat.2328}.  Parameters in bold are the stellar parameters used in priors for our global analysis presented in \autoref{Data_analys}.
	\label{stellarpar}}
\end{table*}

\section{Stellar characterisation} \label{stellar_carac}

\subsection{SED analysis} \label{SED_analysis}

To determine the basic stellar parameters, we performed an analysis of the broadband spectral energy distribution (SED) of TOI-2084 and TOI-4184 together with the {\it Gaia\/} EDR3 parallax \citep[with no systematic offset applied; see, e.g.,][]{Stassun_2021ApJ}, in order to determine an empirical measurement of the stellar radius, following the procedures described in \citet{Stassun_2016AJ,Stassun_2017AJ,Stassun_2018ApJ}. We pulled the  $JHK_S$ magnitudes from {\it 2MASS}, the W1--W3 magnitudes from {\it WISE}, the $G_{\rm BP}$ and $ G_{\rm RP}$ magnitudes from {\it Gaia}, and the $grizy$ magnitudes from {\it Pan-STARRS}. Together, the available photometry spans the full stellar SED over the wavelength range 0.4--10~$\mu$m (see \autoref{SED_plots}). We also estimated the stellar mass according to the empirical $M_K$ based relations of \cite{Mann:2019}.  Deduced stellar parameters of TOI-2084 and TOI-4184 are presented in \autoref{stellarpar}.

\subsection{Spectroscopic analysis}

In addition to the SED analysis, we also compared the Shane/Kast optical spectrum of TOI-2084 to the SDSS M dwarf templates of \citet{2007AJ....133..531B} and found the best match to the M2 template (Figure~\ref{fig:stellar_spectra_OPT}).
The spectral index classification relations of \citet{2003AJ....125.1598L} confirm this classification. We see no evidence of H$\alpha$ emission (equivalent width limit of $<$1.0~{\AA}), indicating an age greater than $\sim$1.2~Gyr \citep{2008AJ....135..785W}. We also measured the $\zeta$ index from TiO and CaH features \citep{2007ApJ...669.1235L,2013AJ....145...52M}, finding $\zeta$ = 0.893$\pm$0.005, consistent with a metallicity of [Fe/H] = $-$0.13$\pm$0.20 based on the calibration of \citet{2013AJ....145...52M}.

For TOI-4184, we also analyzed its \textit{Magellan}/FIRE spectrum using the SpeX Prism Library Analysis Toolkit (SPLAT, \citealt{Burgasser2017}).
By comparing the spectrum to NIR spectral standards defined in \citet{Kirkpatrick2010}, we find the closest match to the M5.0 standard, although the M6.0 standard provides only a marginally poorer match (\autoref{fig:stellar_spectra_NIR}).
Thus, we adopt a spectral type of M5.5$\pm$0.5 for TOI-4184.
We also estimated the metallicity of TOI-4184 from the \textit{Magellan}/FIRE spectrum from the equivalent widths of K-band Na\,\textsc{i} and Ca\,\textsc{i} doublets and the H2O--K2 index \citep{Rojas-Ayala2012}, and used the empirical relation between these observables and stellar metallicity \citep{Mann2014} to estimate [Fe/H].
Following \cite{Delrez_2022_A&A}, we calculated the uncertainty of our estimate using a Monte Carlo approach.
Adding in quadrature the systematic uncertainty of the relation (0.07), we obtained our final estimate of $\mathrm{[Fe/H]} = -0.27 \pm 0.09$, indicating that TOI-4184 is a moderately metal-poor star.

\subsection{The Wide Companion to TOI-2084}

 TOI-2084 has a wide
stellar companion, 2MASS~J17170042+7244364 (hereafter TOI-2084B; $G$=20.7, $J$=16.1), separated by 12$\farcs$2 ($\sim$1400~au) at position angle 191$\degr$ east of north. Both sources are in Gaia DR3 and share a common parallax and proper motion.
The Shane/Kast optical spectrum of TOI-2084B is shown in Figure~\ref{fig:stellar_spectra_OPT}, and is an excellent match to the M8 dwarf template from \citet{2007AJ....133..531B}. This classification is confirmed by the spectral index classification relations of \citet{2003AJ....125.1598L}. We see no evidence of H$\alpha$ emission from this companion, although the noise is considerable in the 6563~{\AA} region. Similarly, we are unable to reliably measure a $\zeta$ index from these data, although the close match to the dwarf template suggests a near-solar metallicity similar to TOI-2084.
There are several known planetary systems orbiting stars in low-mass multiples, including the M4+M4.5 binary TOI-1452 and TOI-1760 \citep{2022AJ....164...96C} and the early-M triple system LTT~1445 \citep{2019AJ....158..152W}.
TOI-2048 and TOI-2048B have an unusually wide separation among low-mass planet hosts in binary systems, although there are examples of such systems among more massive stellar binaries \citep{2017A&A...608A.116C}.

\section{Planet validation} \label{Validate_planet}

\subsection{\emph{TESS} data validation}

  TOI-4184 (TIC 394357918) was observed in sectors 1, 28, and 39 of \emph{TESS} with 2-min cadence. The Science Processing Operations Center (SPOC, \cite{SPOC_Jenkins_2016SPIE}) extracted the photometry of sectors 1 and 28 and performed a transit search (\cite{jenkins2002,jenkins2010}, which yielded the candidate with a 4.902\,day period at a signal-to-noise ratio (S/N) of 11. The SPOC Data Validation (DV) report \cite{Twicken2018,Li2019} was reviewed by the \emph{TESS} Object of Interest (TOI) vetting team on May 27 2021 and TOI-4184 was released on July 8 2021 \citep{guerrero2021}. A second Data Validation Report was issued on the July 24 2021. The transit depth found was 10.7~$\pm$~0.9~ppt, corresponding to a planet radius of 2.6~$\pm$~0.4 R$_\oplus$, and with a period of 4.90199~$\pm$~0.00001~days. The odd/even comparison of the depths agreed to 2.28$\sigma$. Given the large pixel scale of 21~arcsecs, two neighboring stars were fully or partially included in the \emph{TESS} aperture, as seen in \autoref{Target_pixel}, although TOI-4184 was identified as the likely source of the events. The difference imaging centroid test performed for sectors 1 to 39 constrains the location of the target star's catalog location to 5.5 $\pm$ 3.3 arcsec. All additional validation tests, including centroid offset, bootstrap, and ghost tests, were passed. 

 TOI-2084 (TIC 441738827) was observed in sectors 16, 19--23, 25--26, and 48--60 with 2-min cadence by \emph{TESS}. The SPOC pipeline produced the first DV report on May 6 2020 using extracted photometry of sectors 16 and 19--23. Two candidates were identified: .01 has a period of 6.078 days and a depth of 2.760~$\pm$~258 ppt at an S/N of 11.2, and .02 a period of 8.149 days and a depth of 3.313~$\pm$~327 ppt at an S/N of 10.8. The report was reviewed by the TOI vetting team and the candidates were released on July 15 2020. A second DV report was issued on August 7, 2020 from the SPOC pipeline which included sectors up to 26 of 2-min cadence data. The first candidate was found to have a period of 6.07830~$\pm$~0.00010 days, a transit depth of 2.8~$\pm$~0.2 ppt with an S/N of 12.7, and a planetary radius of 2.6~$\pm$~0.7 R$_\oplus$. Silimarly, the second candidate was found to have a period of 8.14903~$\pm$~0.00018 days, a transit depth of 2.8~$\pm$~0.2 ppt with an S/N of 11.8, and a planetary radius of 2.6~$\pm$~0.6 R$_\oplus$. The odd/even phase-folded transits were compared and agreed to $1.45 \sigma$ and $0.96 \sigma$ for the .01 and .02 candidates, respectively. As for TOI-4184, one nearby star is contaminating the aperture, but the event was limited to be on target for the .01 candidate and likely on target for .02. In addition, the DV report a difference imaging centroid test result that locates the catalog position of the target star to within 2.0$\pm$3.0 arcsec. The rest of the validation tests were passed. The \emph{TESS} full-frames images were also processed by the Quick Look Pipeline (QLP, \cite{QLP}), which extracted the photometry of TOI-2084 in sectors 15--16, 18--25, and 47--52, and confirmed the signal found in 2-min cadence data. 

\subsection{Archival imaging}
\begin{figure}
	\centering
	\includegraphics[scale=0.22]{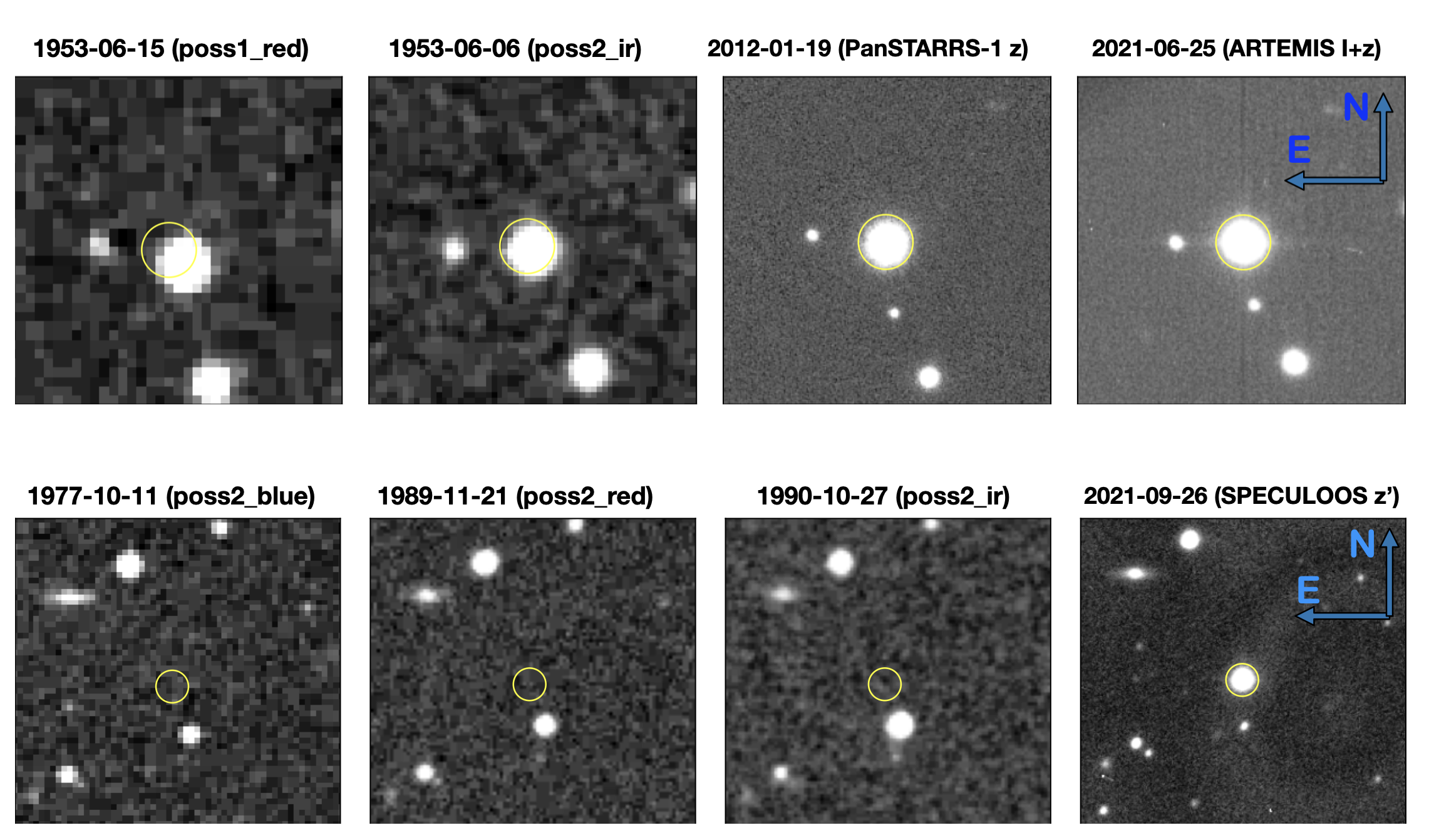}
	\caption{Field images cropped on a 1'$\times$1' region around  TOI-2084 (top row of images) and TOI-4184 (bottom row). The current position of the target stars is shown with the yellow circle. {\it Top row}, from left to right: 1953 red image from POSS I/DSS, 1953 infrared image from POSS II/DSS2, 2012 \textit{z'} image from PanSTARRS1, and 2021 \textit{I+z} image from SPECULOOS-North. {\it Bottom row}, from left to right: 1977 blue image from POSS II/DSS2, 1989 red image from POSS II/DSS2, 1990 infrared image from POSS II/DSS2, and 2021 image \textit{z'} from SPECULOOS-South.}
	\label{archival_images}
\end{figure}

We obtained archival images of TOI-2084 and TOI-4184 in order to discard the case of a background unresolved companion producing the transit signals. Whether an eclipsing binary, a planetary candidate orbiting a background star, or simply an unaccounted background star, any of these scenarios would skew the stellar and planetary parameters obtained from our global analysis. TOI-2084 has a relatively low proper motion of 60.24\,mas\,yr$^{-1}$, which makes it challenging to assess the background of the star's current position. We obtained images from POSS I/DSS (\cite{1963POSS-I}), POSS II/DSS (\cite{1996DSS_POSS-II}) and PanSTARRS1 (\cite{2016_Pan-STARRS1}) in the red, infrared and $z$ bands, respectively, and spanning 68 years, as shown in \autoref{archival_images}. TOI-2084 has moved by $> 4\arcsec$ between the 1953 image and the 2021 image. Given the pixel scale of 1--1.7\arcsec, it is impossible to rule out a background star from this diagnostic alone, though it is unlikely since we ruled out any close companion star at a minimum angular separation of $0.1 \arcsec$ (see Section \ref{High_Reso_Imag}).
We also compared images centered on TOI-4184 from POSS II/DSS  in the blue, red and infrared bands taken in 1977, 1989, and 1990, respectively. Because of its high proper motion of 183.87\,mas\,yr$^{-1}$, TOI-4184 has moved by $> 8\arcsec$ in the 44 years spanning the observations. This allows us to confirm the lack of background contaminant in the line-of-sight brighter than a limiting magnitude of $\geq 20$. 

\subsection{Follow-up photometric validation} \label{val_follo_photometric}
\begin{figure*}
	\centering
	\includegraphics[scale=0.35]{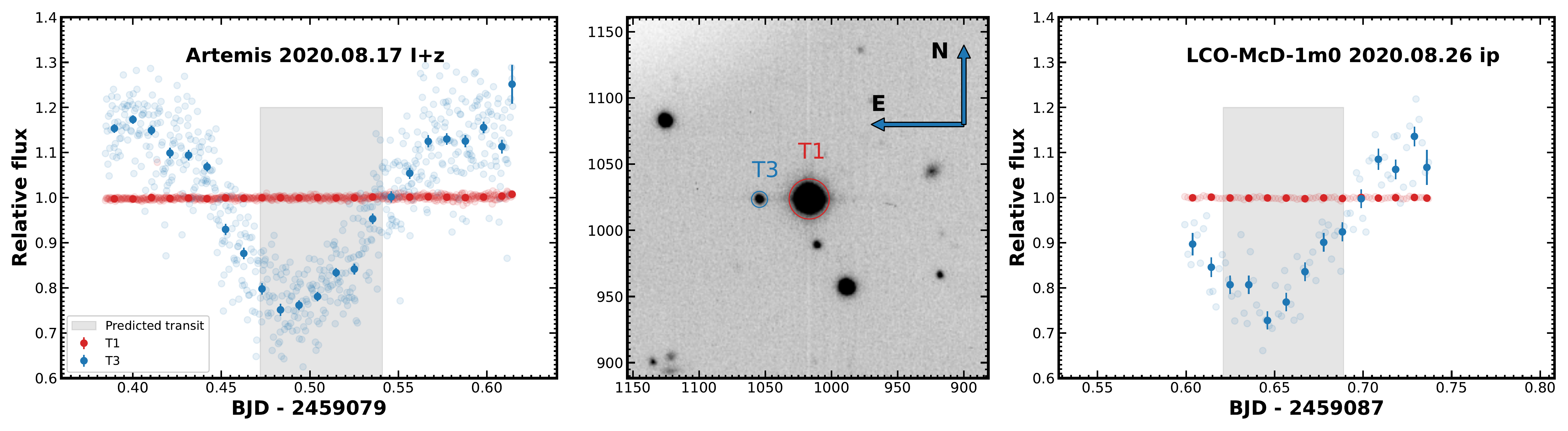}
	\caption{TOI-2084 light curves obtained with ground-based facilities. {\it Left panel:} light curve obtained with SPECULOOS-North in the \textit{I+z} filter on UTC August 17 2020. {\it Middle panel:} TOI-2084 field-of-view with nearby stars.  The wide co-moving companion TOI~2084B is directly south. {\it Right panel:} light curve obtained with LCO-McD in the Sloan-$i'$ filter on UTC August 26 2020. Red and blue data points show the target (T1) and nearby star (T3) light curves, respectively. During the expected transit of TOI-2084.02, we twice detected a deep eclipse ($\approx 400$~ppt) on the nearby star TIC 1271317080 ($\Delta T = 4.98$) at 12.9" from the target, labeled T3.}
	\label{NEB_check}
\end{figure*}

\begin{figure}
	\centering
	\includegraphics[scale=0.4]{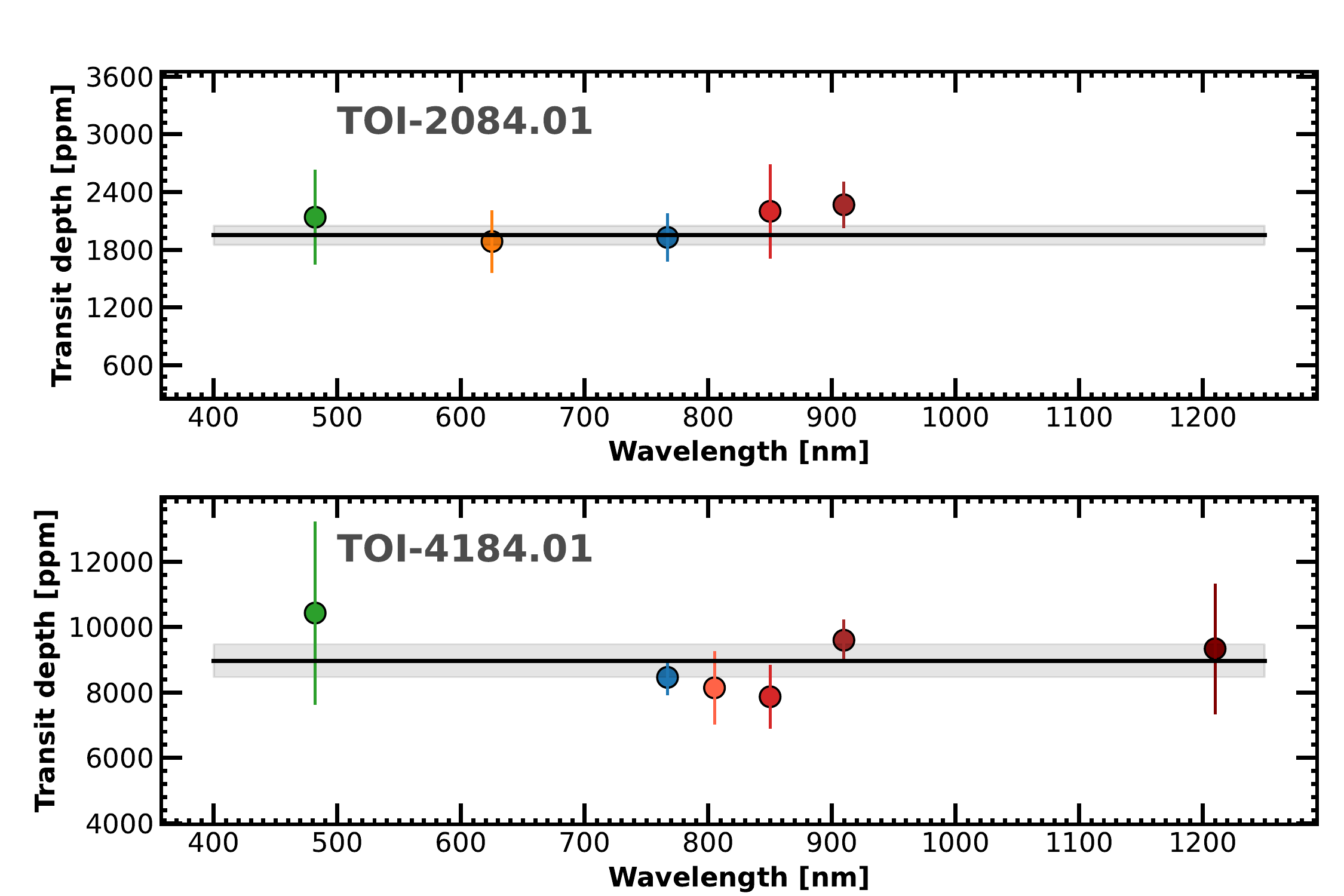}
	\caption{Transit depths measured for multi-band photometric follow-up of TOI-2084.01 (top panel) and TOI-4184.01 (bottom panel). Solid horizontal lines correspond to the \emph{TESS} measured transit depth.}
	\label{Depth_transit}
\end{figure}

Photometric follow-up using ground-based facilities has two objectives: identify the source of the transit event and assess if the transit depth is wavelength dependent. The presence of contaminating stars in the \emph{TESS} aperture was noted for both TOI-4184.01 and TOI-2084.01 in the \emph{TESS} data validation reports. The closest neighboring stars are respectively TIC 650071720 at $11.5\arcsec$ with a $\Delta$Tmag of 4.45, and TIC 441738830 at $12.4\arcsec$ with a $\Delta$Tmag of 6.15. We reached aperture sizes of a few arcseconds using ground-based facilities, which allowed us to confirm the transit events are on the expected stars for TOI-4184.01 and TOI-2084.01. In the case of TOI-2084.02, twice at the expected transit times we detected a deep eclipse on the nearby star TIC\,1271317080 ($\Delta T = 4.98$) at 12.9" from the target, labeled T3 in \autoref{NEB_check}.
Thus, we rule out TOI-2084.02 as a false positive and do not consider it further.
We collected photometric data for TOI-2084.01 in various bands (\textit{I+z}, \textit{zs}, \textit{i'}, \textit{r'}, \textit{g'}), spanning the 400--1100\,nm wavelength ranges. We measured a matching transit depth within $1\sigma$ in all bands. Similarly, we obtained data for TOI-4184.01 in the \textit{I+z}, \textit{zs}, \textit{Ic}, \textit{JJ}, \textit{g'} bands, covering a range between 400--1210\,nm where the transits depths also agree within $1\sigma$. The transit depths measured in different wavelengths for TOI-2084\,b and TOI-4184\,b are presented in \autoref{Depth_transit}, \autoref{toi2084_param} and \autoref{toi4184_param}.

\subsection{Statistical validation}
 
To calculate the false positive probability (FPP) for TOI-2084.01 and TOI-4184.01, we used the Tool for Rating Interesting Candidate Exoplanets and Reliability Analysis of Transits Originating from Proximate
Stars \citep[\triceratops ;][]{Giacalone_2021AJ}. This Bayesian tool incorporates prior knowledge of the target star, planet occurrence rates, and stellar multiplicity to calculate the probability that a given transit signal is due to a transiting planet or another astrophysical source. The criteria for statistical validation of a planetary candidate is stated as 
$FPP\footnote{FPP: false positive probability} < 0.01$ and $NFPP\footnote{NFPP: nearby false positive probability} < 0.001$, which is the sum of probabilities for all false positive scenarios. 
We ran TRICERATOPS on the \emph{TESS} light curves including the contrast curve obtained with Gemini/Alopeke and Gemini/Zorro for both stars, TOI-2084 and TOI-4184. We found $FPP = 0.0005$  and $FPP = 0.0001$  for TOI-2084\,b and TOI-4184\,b, respectively. Because triceratops determines that no nearby stars are capable of being sources of astrophysical false positives, we find $NFPP = 0$ for both candidates (TOI-2084.01 and TOI-4184.01). Based on these results, we consider two planets to be validated.
TOI-2084.02 was rejected as a nearby eclipsing binary (NEB) based on ground-based photometric follow-up (see previous Section \ref{val_follo_photometric}).

\section{Photometric data modelling} \label{Data_analys}

\begin{table*}
 \begin{center}
 {\renewcommand{\arraystretch}{1.4}
 \begin{tabular}{l c c l c c cc ccc}
 \toprule
Planet & Telescope & Filter   & Baseline model & $\beta_{w}$ & $\beta_{r}$ & $CF$\\ 
 \hline
 TOI-2084.01 & Artemis-1.0m (obs1) & $I+z$ & Time$^1$ & 1.00 &  1.02 &  1.03 \\
 TOI-2084.01 & TRAPPIST-N-0.6m & $I+z$  & Time$^1$, Fwhm$^1$ & 1.07 &  1.09 &  1.17 \\
 TOI-2084.01 & MuSCAT3-2.0m & Sloan-$g'$  & Time$^1$, Airmass$^1$ & 0.90 &  1.60 &  1.44 \\
 TOI-2084.01 & MuSCAT3-2.0m & Sloan-$r'$  & Time$^1$ & 1.04 &  1.01 &  1.05\\
 TOI-2084.01 & MuSCAT3-2.0m & Sloan-$i'$  & Time$^1$ & 1.12 &  1.37 &  1.54\\
 TOI-2084.01 & MuSCAT3-2.0m & Pan-STARRS-$zs$  & Time$^1$ & 1.15 &  1.00 & 1.15\\
 TOI-2084.01 & Artemis-1.0m (obs2) & $I+z$ & Time$^2$& 1.04 &  1.45 &  1.51 \\
 TOI-2084.01 & TRAPPIST-N-0.6m & $I+z$  & Time$^1$, Flip & 1.01 &  1.14 &  1.16\\
 TOI-2084.01 & SAINT-EX-1.0m  & Sloan-$r'$ & Time$^1$ & 0.66 &  1.44 &  9.48 \\
 \hline
 TOI-4184.01 & LCO-CTIO-1.0m & Sloan-$i'$ & Time$^2$ & 0.68 &  1.10 & 0.75 \\
 TOI-4184.01 & TRAPPIST-S-0.6m & $I+z$ & Time$^1$, Fwhm$^1$ &  0.49 & 1.32 & 0.64\\
 TOI-4184.01 & Danish-1.54m& $I_C$ & Time$^2$ & 0.97 & 1.08 & 1.05\\
 TOI-4184.01 & ExTrA-0.6m & $1.21 \mu m$ &  Time$^1$ & 1.01 & 1.16 &  1.17 \\
 TOI-4184.01 & Danish-1.54m & $I_C$ & Time$^2$ & 0.98 & 1.42 &  1.40\\
 TOI-4184.01 & TRAPPIST-S-0.6m & $I+z$ & Time$^1$, Airmass$^1$ & 0.73 &  1.44 &  1.04 \\
 TOI-4184.01 & LCO-CTIO-1.0m & Sloan-$i'$ & Time$^2$ & 0.80 & 1.15 &  9.13 \\
 TOI-4184.01 & SPECULOOS-S-1.0m & Sloan-$z'$ & Time$^1$ & 0.62 & 1.01 &  0.63 \\
 TOI-4184.01 & LCO-SAAO-1.0m & Sloan-$i'$ & Time$^2$ & 1.21 &  1.47 &  1.77 \\
 TOI-4184.01 & LCO-SAAO-1.0m & Sloan-$g'$ & Time$^2$ &  0.86 &  1.11 &  0.96\\
 \hline
 \end{tabular}}
 \caption{MCMC analysis parameters. For each transit light curve  selected baseline-function (based on the BIC), deduced values of $\beta_{\rm w}$, $\beta_{\rm r}$ and the coefficient correction  $CF =  \beta_{\rm w} \times \beta_{r}$.}
 \label{baseline_table_1}
 \end{center}
\end{table*}

We performed a joint fit of all observed light curves by \emph{TESS} and ground-based telescopes described in \autoref{photometric_observation}, using the Metropolis-Hastings \citep{Metropolis_1953,Hastings_1970} algorithm implemented in the updated version of MCMC (Markov-chain Monte Carlo) code described in \cite{Gillon2012}. The transit light curves are modeled using the quadratic limb-darkening model of \cite{Mandel2002}, multiplied by a baseline model in order to correct for several external effects related to systematic variations (time, airmass, background, full-width half-maximum, and position on the detector). The baseline model was selected based on minimizing the Bayesian information criterion (BIC) described in \cite{schwarz1978}. 
 \autoref{baseline_table_1} shows for each transit light curve the selected baseline model based on the BIC, and correction factor $CF = \beta_{w} \times \beta_{r}$ to rescale the photometric errors, where $\beta_{w}$ and $\beta_{r}$ are white and red noises, respectively (see \cite{Gillon2012} for more details). 
  TRAPPIST-South and TRAPPIST-North telescopes are equipped with German equatorial mounts that have to rotate $180^\circ$ when the meridian flip is reached. This movement results the stellar images in different position on the detector before and after the flip. The normalization offset is included as jump parameter in our global MCMC analysis. The transit light curve observed with TRAPPIST-South on UTC September 20 2021  contains a meridian flip at BJD = 2459478.8226 (see \autoref{obs_table}), which is accounted during the global analysis.

The jump parameters sampled by the MCMC for each system were: 
\begin{itemize}
    \item $T_0$: the transit timing;
    \item $W$: the transit duration (duration between the contacts 1 and 4);
    \item $R_p^2/R_\star^2$: the transit depth, where $R_p$ is the planet radius and $R_\star$ is the star radius;
    \item $P$: the orbital period of the planet;
    \item $b = a \cos(i_p)/R_\star$: the impact parameter in case of the circular orbit, where $i_p$ is the planetary orbital inclination and $a_p$ is the semi-major axis of the orbit;
    \item $\sqrt{e}\cos(\omega)$, were $\omega$ is the argument of periastron and $e$ is the orbital eccentricity
    \item the combination $q_1 = (u_1 + u_2)^2$ and $q_2 = 0.5u_1(u_1 + u_2)^{-1}$ \citep{Kipping_2013MNRAS.435.2152K}, were $u_1$ and $u_2$ are the quadratic limb-darkening  coefficients, which are calculated from \cite{Claret_2012AA};
    \item and the stellar metallicity $[Fe/H]$, the effective temperature ($T_{\rm eff}$), log of the stellar density ($\log(\rho_\star)$), and log of the stellar mass ($\log(M_\star$)).
\end{itemize}

For each star, we applied a Gaussian prior distribution on the stellar parameters obtained from SED and spectroscopy (which are $R_\star$, $M_\star$, $[Fe/H]$, $\log g_\star$ and $T_{\rm eff}$).
For each system, we performed two MCMC analyses, the first assuming a circular orbit, and the second assuming an eccentric orbit. The results are compatible with a circular orbit based on the Bayes factor $BC = \exp{(-\Delta BIC/2)}$. The eccentric solutions give $e {\sim}0.2^{+0.3}_{-0.2}$ for TOI-2084.01 and $e {\sim}0.1^{+0.2}_{-0.1}$ for TOI-4184.01. 

For each transit light curve, a preliminary analysis composed of one Markov chain of $10^5$ steps was performed in order to calculate the correction factor $CF$. Then a global MCMC analysis of three Markov chains of $10^5$ steps was performed to derive the stellar and planetary physical parameters. The convergence of each Markov chain was checked using the statistical test of \cite{Gelman1992}. Derived parameters of TOI-2084 and TOI-4184 are presented in Tables \ref{stellarpar}, \ref{toi2084_param} and \ref{toi4184_param}.

\section{Planet searches using the TESS photometry} \label{search}

In this section, we searched for additional planetary candidates that might remain unnoticed by SPOC and the QLP due to their detection thresholds. To this end we used our custom pipeline \sherlock\footnote{\sherlock (\textbf{S}earching for \textbf{H}ints of \textbf{E}xoplanets f\textbf{R}om \textbf{L}ightcurves \textbf{O}f spa\textbf{C}e-based see\textbf{K}ers) code is fully available on GitHub: \url{https://github.com/franpoz/SHERLOCK}}, originally presented by \cite{pozuelos2020} and \cite{Demory_AA_SAINTEX_2020}, and used in several studies \citep[see, e.g.,][]{Wells_2021A&A_TOI2406b,vangrootel2021,Schanche_2022A&A}. 

\sherlock allows the user to explore TESS data to recover known planets, alerted candidates, and search for new periodic signals, which may hint at the existence of extra transiting planets.
In short, the pipeline has six modules to (1) download and prepare the light curves from the MAST using the \lightkurve \citep{lightkurve}, (2) search for planetary candidates through the \tls \citep{tls}, (3) perform a semi-automatic vetting of the interesting signals, (4) compute a statistical validation using the \triceratops \citep{Giacalone_2021AJ}, (5) model the signals to refine their ephemerides employing the \allesfitter package \citep{allesfitter}, and (6) compute observational windows from ground-based observatories to trigger a follow-up campaign. We refer the reader to \cite{Delrez_2022_A&A} and \cite{Pozuelos_2023A&A} for recent \sherlock applications and further details. 

For  TOI-4184, we searched for extra planets analyzing the three available sectors (1, 28, and 39) together, exploring orbital periods from 0.3 to 30~d. For TOI-2084.01, we conducted two independent searches: 1) corresponding to the nominal mission, that is, 8 sectors from 16 to 26, and 2) corresponding to the extended mission, that is, 13 sectors from 48 to 60 (see ~\autoref{TOI_2084_4184_TESS_LC}). In both searches, we explored the orbital periods from 0.3 to 50. The motivation to follow this strategy is twofold. On the one hand, the high computational cost of exploring at the same time 21 sectors, while adding many sectors might hint at the presence of very long orbital periods ($>50$ days), the transit probabilities rapidly decrease for such scenarios. On the other hand, this strategy allows us to compare any finding in the nominal mission with the extended mission, providing an extra vetting step for the signals' credibility. 

We successfully recovered the TOIs released by SPOC, the TOI-4184.01 with an orbital period of 4.90 days and TOI-2084.01 with an orbital period of 6.08 days. In the subsequent runs performed by \sherlock, we did not find any other signal that hinted at the existence of extra transiting planets. In addition to TOI-2084.01, we also recovered a signal corresponding to TOI-2084.02, which was already classified as a false positive using ground-based observations described Section~\ref{val_follo_photometric} and displayed in Figure~\autoref{NEB_check}. Surprisingly, we did not recover the signal with the orbital period issued by TESS, 8.14~days, but its first subharmonic, which corresponds to an orbital period of 4.07~days. Then, we used the two modules implemented in SHERLOCK for vetting and statistical validation of candidates with this signal. On the one hand, using the vetting module, we found that even and odd transits yielded different transit depths; $\sim$2.3 and $\sim$1.1 ppt for even and odd transits, respectively. This indicated that our detection algorithm was confusing the secondary eclipse as the primary and yielding half of the real orbital period, which confirmed that the real orbital period is 8.14~days. On the other hand, the validation module found that its FFP is $\sim$0.26 and NFPP is $\sim$0.1. According to \cite{Giacalone_2021AJ}, these values place this candidate in the false positive area in the NFPP-FPP plane. Hence, these analyses agreed with the eclipsing binary nature of this signal.

\section{Results and discussion} \label{Result_discuss}

We presented the validation and discovery of TOI-2084\,b and TOI-4184\,b by the \emph{TESS} mission (see phase-folded light curves in \autoref{plot_lightcurves_TESS} and individual transits in \autoref{TOI_2084_4184_TESS_LC}), which were confirmed through follow-up photometric measurements collected by SPECULOOS-South/North, SAINT-EX, TRAPPIST-South/North, MuSCAT3, LCOGT, Danish and ExTrA telescopes (see phase-folded light curves in \autoref{plot_lightcurves_GB}).  
The host stars are characterized by combining optical spectrum obtained by Shane/Kast and Magellan/FIRE, SED and stellar evolutionary models. Then, we performed a global analysis of space \emph{TESS} and ground-based photometric data to derive the stellar and planetary physical parameters for each system.  
\autoref{stellarpar} shows the astrometry, photometry, and spectroscopy stellar properties of TOI-2084 and TOI-4184.
Derived stellar and planetary physical parameters from our global analysis are shown in \autoref{toi2084_param} and \autoref{toi4184_param}. \autoref{plot_periodogram} shows the periodogram for the system. Both planets are well detected in \emph{TESS} data. \autoref{corner_TOI2084} and \autoref{corner_TOI4184} show the parameters posterior distributions for each system.

\subsection{TOI-2084\,b and TOI-4184\,b}

TOI-2084 is a $K_{\rm mag} = 11.15$ M2-type star with an effective temperature of $T_{\rm} = 3553 \pm 50$K, a surface gravity of $\log g_\star = 4.75 \pm 0.05$~dex, a mass of $M_\star = 0.49 \pm 0.03~M_\odot$ and a radius $R_\odot = 0.475 \pm 0.016~R_\odot$ (derived from SED analysis including Gaia EDR3 parallax) and a metallicity of $[Fe/H] = -0.13 \pm 0.20$ (from Shane/Kast  spectrum).
 It has a wide ($\sim$1400~au) M8 co-moving companion, with a likely mass of 0.1~$M_\odot$.
TOI-2084\,b is a sub-Neptune-sized planet orbiting around  the host primary star every 
6.08~days, which has a radius of $R_p = 2.47 \pm 0.13~R_\oplus$, an equilibrium temperature of $T_{\rm eq} = 527 \pm 8$~K, an incident flux of $S_p = 12.8 \pm 0.8$ times that of Earth. We find that TOI-2084\,b has a predicted mass of $M_p = 6.74^{+5.31}_{-2.81}~M_\oplus$ using the \cite{Chen_Kipping_2017ApJ} relationship. 

TOI-4184 is a $K_{\rm mag} = 11.86$ M5.5$\pm$0.5 metal-poor star with a metallicity of $[Fe/H] = -0.27 \pm 0.09$~dex (from the Magellan/FIRE spectrum), an effective temperature of $T_{\rm eff} = 3225 \pm 75$~K, a surface gravity of $\log g_\star = 5.01 \pm 0.04$~dex, a mass of $M_\star = 0.240 \pm 0.012~M_\odot$ and a radius $R_\odot = 0.242 \pm 0.013~R_\odot$.
TOI-4184\,b is a sub-Neptune-sized planet that completes its orbit around its host star in  4.9~days, has a radius of $R_p = 2.43 \pm 0.21~R_\oplus$, an irradiation of $S_p = 4.8 \pm 0.4$ Earth irradiation, and an equilibrium temperature of $T_{\rm eq} = 412 \pm 8$~K. We used the \cite{Chen_Kipping_2017ApJ} relationship to predict the plausible mass of TOI-4184\,b, which is $M_p = 6.60^{+5.20}_{-2.75}~M_\oplus$.

\autoref{Desert_Neptune} shows the boundaries of the sub-Neptune-desert region determined by \cite{Mazeh_2016AandA_Jovian_desert}. TOI-2084\,b and TOI-4184\,b are placed at the edge of  the sub-Jovian desert in the radius-period plane. Combining the infrared brightness of the host star and predicted semi-amplitude of the radial-velocity ($K_{\rm TOI-2084b} = 3.8^{+3.0}_{-1.6}$~m/s for TOI-2084\,b and $K_{\rm TOI-4184b} = 6.4^{+5.0}_{-2.7}$~m/s for TOI-4184\,b) using \cite{Chen_Kipping_2017ApJ}'s relationship,  make TOI-2084\,b and TOI-4184\,b good targets for radial velocity follow-up with high-resolution spectrographs (e.g., CARMENES, \cite{Quirrenbach_2014IAUS}, ESO-VLT-8.0m/ESPRESSO, \cite{Pepe_2021A&A}, Gemini-North-8.0m/MAROON-X, \cite{MAROON-X2021} and ESO-3.6m/NIRPS, \cite{Bouchy_2021spc_confE}) to constrain the planetary masses, bulk densities and other orbital parameters.

\subsection{Characterization prospects}
Super-Earths and sub-Neptunes are amongst the most abundant type of exoplanets. Yet their formation, atmospheric composition, and interior structure are not well understood, as a variety of compositions can match the average density of these planets. TOI-2084\,b and TOI-4184\,b are part of this mysterious population. The small size and proximity of the host stars as well as their brightness in the infrared make them amenable to be further observed by most JWST modes for studying atmospheric compositions.
\begin{figure}
	\centering
	\includegraphics[scale=0.3]{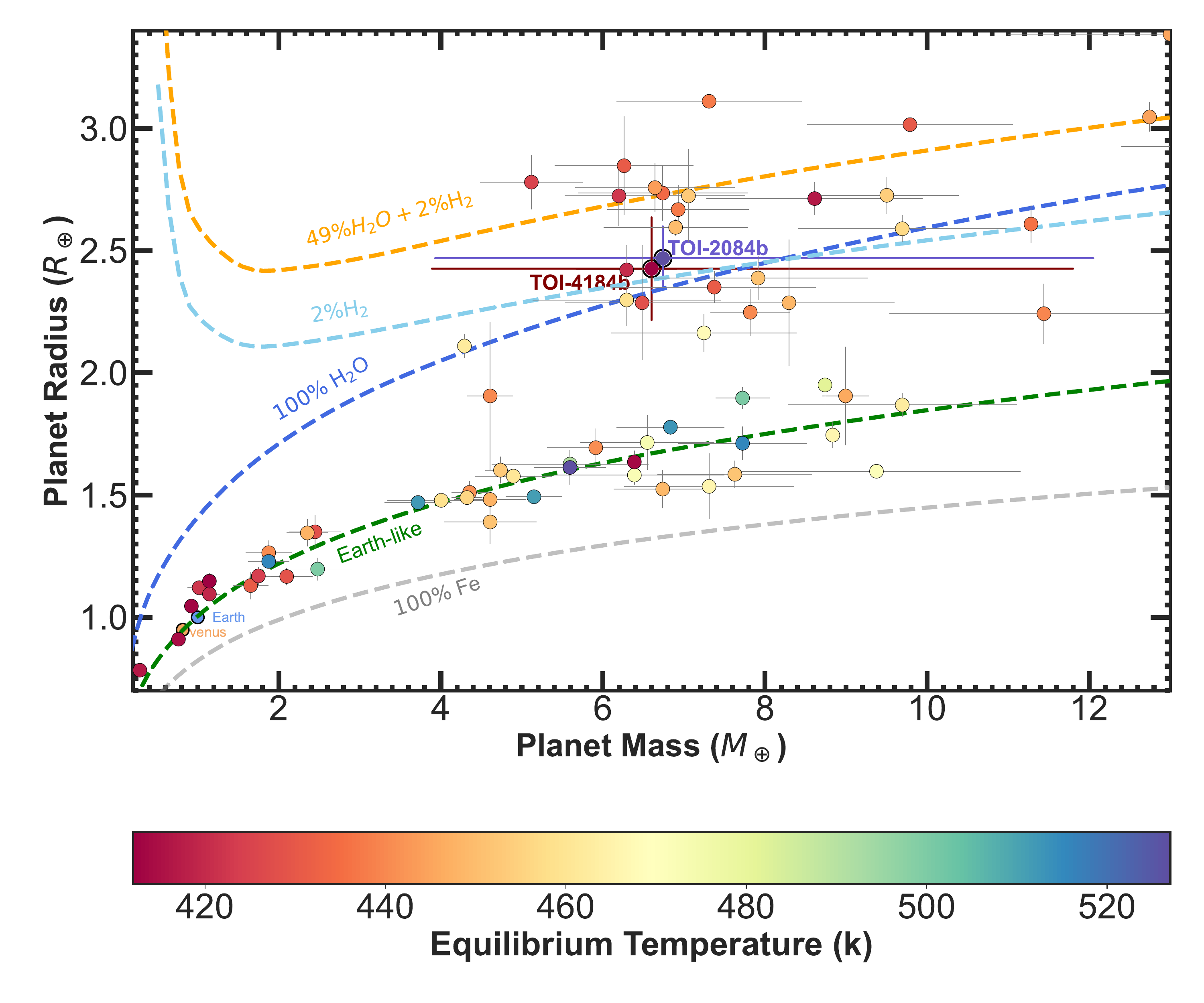}
	\caption{Mass-radius diagram of exoplanets with mass-radius measurements better than 25$\%$ from \href{http://www.astro.keele.ac.uk/jkt/tepcat/}{TEPCat} and for our candidates, color-coded by their equilibrium temperature. Two-layer models from \citet{zeng} are displayed with different lines and colors. "Earth-like" here means a composition of 30$\%$ Fe and 70$\%$ MgSiO$_{3}$.  The 2\% H$_2$ line represents a composition consisting of a 98\% Earth-like rocky core and a 2\% H$_2$ envelope by mass, while the 49\% H$_2$O + 2\% H$_2$ line corresponds to a composition comprising a 49\% Earth-like rocky core, a 49\% H$_2$O layer, and a 2\% H$_2$ envelope by mass. Earth and Venus are identified in this plot as pale blue and orange circles,  respectively.}
	\label{mr}		
\end{figure}

Given the measured properties, we made a first exploratory guess of the planet's composition. We compared their masses and radii with the models from \citet{zeng}, shown in \autoref{mr}. The models predict that TOI-2084\,b and TOI-4184\,b may have low-density volatiles, such as water, an H/He atmosphere, or a combination of both. Below, we further explore these plausible atmospheres and assess the potential for atmospheric characterization of both TOI-2084\,b and TOI-4184\,b planets.\\\\
As a first approximation of the suitability of both planets for atmospheric investigations, we calculate the transmission spectroscopic metric (TSM) from \citet{kem}, which was developed based on simulations with NIRISS.  We estimate the TSMs for TOI-2084\,b and TOI-4184\,b to be $26.7^{+14.7}_{-10.3}$ and $57.7^{+25.7}_{-20.1}$, respectively. With 90 being the threshold for this category of planets, it is worth noting that this metric solely considers the predicted strength of an atmospheric detection when ranking the planets. Having TSM values below the threshold does not necessarily mean that detailed atmospheric studies are impossible or challenging with current facilities. In other words, these metrics do not serve as the sole criterion for determining the best targets for atmospheric studies. \\\\
\begin{figure*}
	\centering
	\includegraphics[width=\columnwidth]{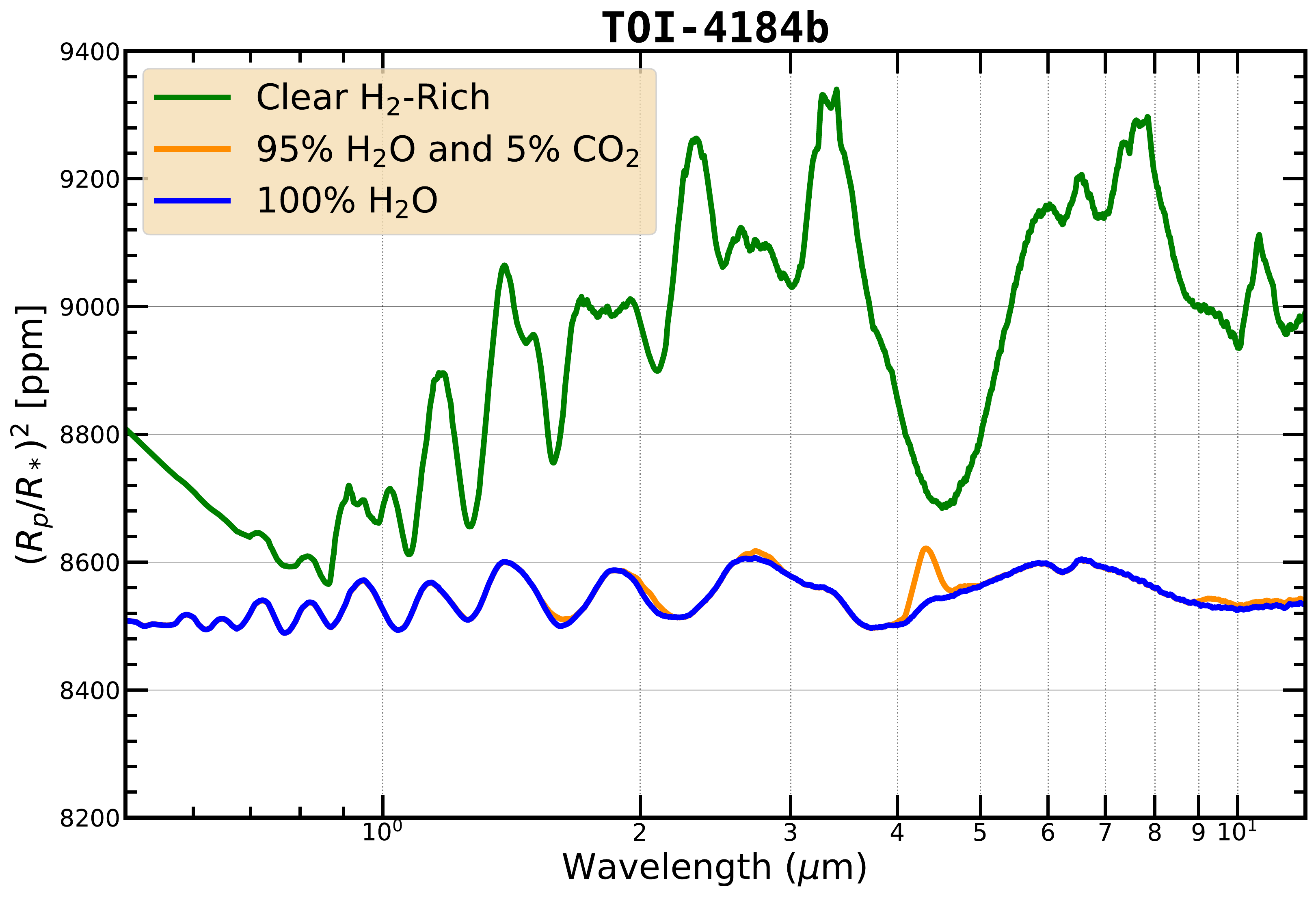} \hfill
    \includegraphics[width=\columnwidth]{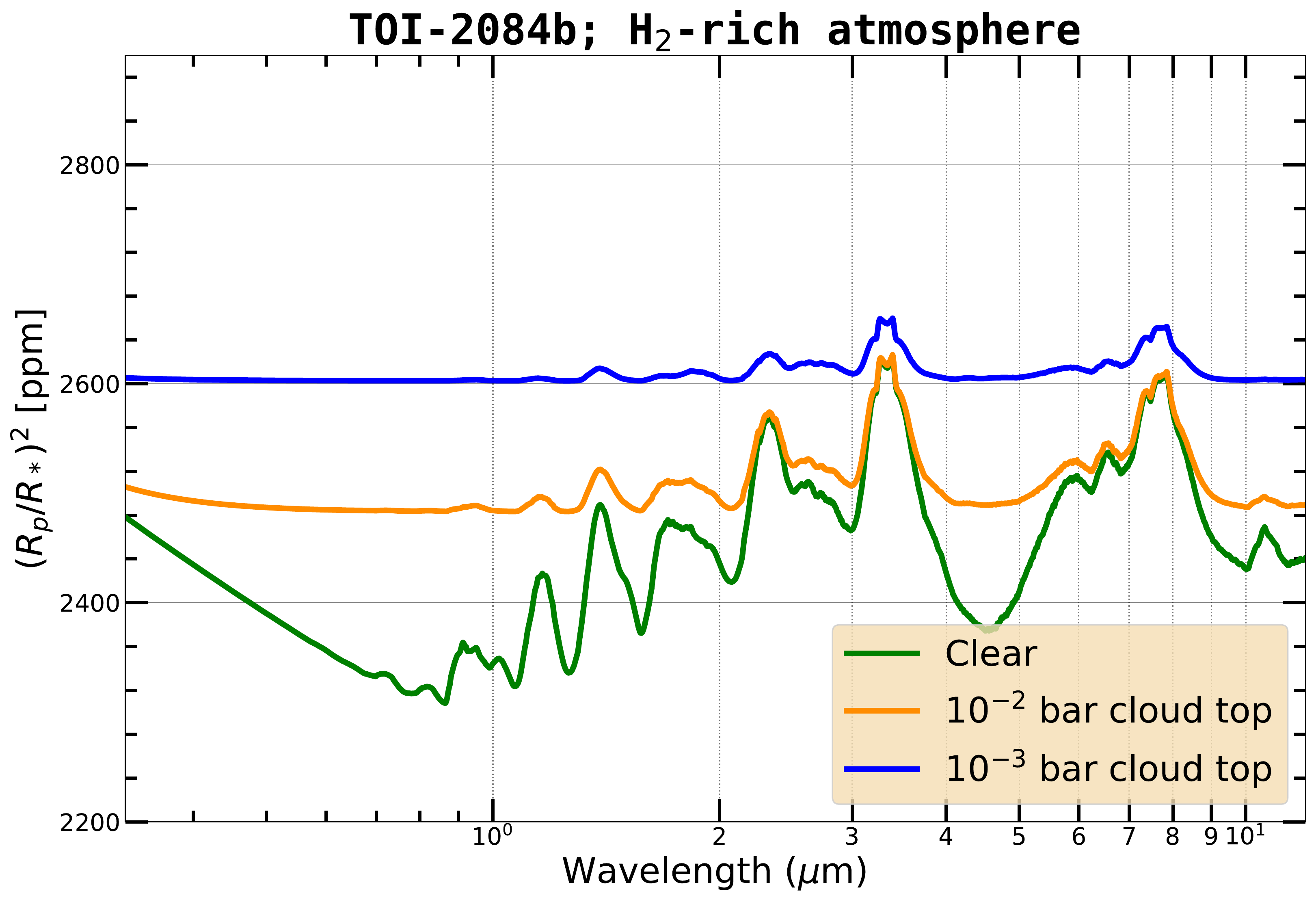}
\caption{Synthetic transmission spectra of TOI-4184b for different atmospheric compositions. Left panel: transmission spectra for cloud-free H$_2$-rich, water-rich (95\% H$_{2}$O and 5\% CO$_{2}$), and pure water atmospheres. Right panel: transmission spectra for H$_2$-rich cloudy atmosphere with cloud layers at different altitudes (10$^{-2}$, 10$^{-3}$, and cloud-free).}
	\label{tr}		
\end{figure*}
To further evaluate the feasibility of characterizing the atmosphere of both planets, we computed synthetic transit spectra from optical to infrared wavelengths (0.5--12\,$\mu$m) at low spectral resolutions for different atmospheric scenarios (cloud-free H$_2$- and cloudy H$_2$- rich, water-rich). We used {\tt petitRADTRANS} \citep{mol} to compute the model transmission spectra, using the stellar parameters from Table \ref{stellarpar} and the planetary parameters from Tables \ref{toi2084_param} \& \ref{toi4184_param}. Our test H$_2$-rich models assume atmospheric chemical equilibrium computed using the FastChem code \citep{fastchem} with isothermal profiles at the equilibrium temperature, solar abundances, collisionally induced absorption (CIA) by H$_2$-H$_2$ and H$_2$-He, and Rayleigh scattering. We include as absorbers H$_{2}$O, CO$_{2}$, CO, CH$_{4}$, NH$_{3}$, C$_{2}$H$_{4}$, and C$_{2}$H$_{2}$. For the water-rich scenarios, we assume that the planets are enveloped in a clear, isothermal water-dominated atmosphere composed of 95\% H$_{2}$O and 5\% CO$_{2}$. The model includes the H$_{2}$O and CO$_{2}$ Rayleigh scattering cross-sections. We also compare it to a pure water planet (100\% H$_{2}$O) with H$_{2}$O Rayleigh scattering. An example of the resulting spectra for TOI-4184b is shown in \autoref{tr}.\\\\
As predicted by earlier studies (e.g., \cite{Greene_2016, molliere2017modeling, Chouqar_2020}), the amplitude of the transmission spectra is highly dependent on the presence and altitude of the cloud layer, and on the average molecular weight of the atmosphere: the higher the average molecular weight of the atmosphere, the lower the scale height, and thus the lower the amplitude of the transit spectroscopy signal. The transmission spectra for the H-rich atmospheres show strong absorption features due to H$_2$O, CH$_4$, and NH$_3$ over the wavelength range 0.5--12\,$\mu$m (see  \autoref{tr}). The spectroscopic modulations of the cloud-free spectra are on the order of 50–350 ppm and 100–700 ppm for TOI-2084\,b and TOI-4184\,b, respectively. The cloudy models present smaller absorption features due to the suppression of contributions from deeper atmospheric layers. The features are essentially muted in the cases with 10$^{-4}$ bar cloud top model for both planets (not shown here). For scenarios discussed above, the mean molecular weight varies from $\mu = 2$~g/mol for an atmosphere dominated by molecular hydrogen to $\mu = 18$~g/mol for atmospheres dominated by heavier molecules like H$_{2}$O which explains the weak spectral features seen in the water-rich atmosphere. Additionally, an atmosphere with 95\% H$_{2}$O and 5\% CO$_{2}$ can be distinguished from an atmosphere with 100\% H$_{2}$O because the transit depth within the CO$_{2}$ band at 4.5\,$\mu$m would be higher relative to the transit depths in the H$_{2}$O bands, as shown in \autoref{tr}.
\begin{figure}
	\centering
	\includegraphics[width=\columnwidth]{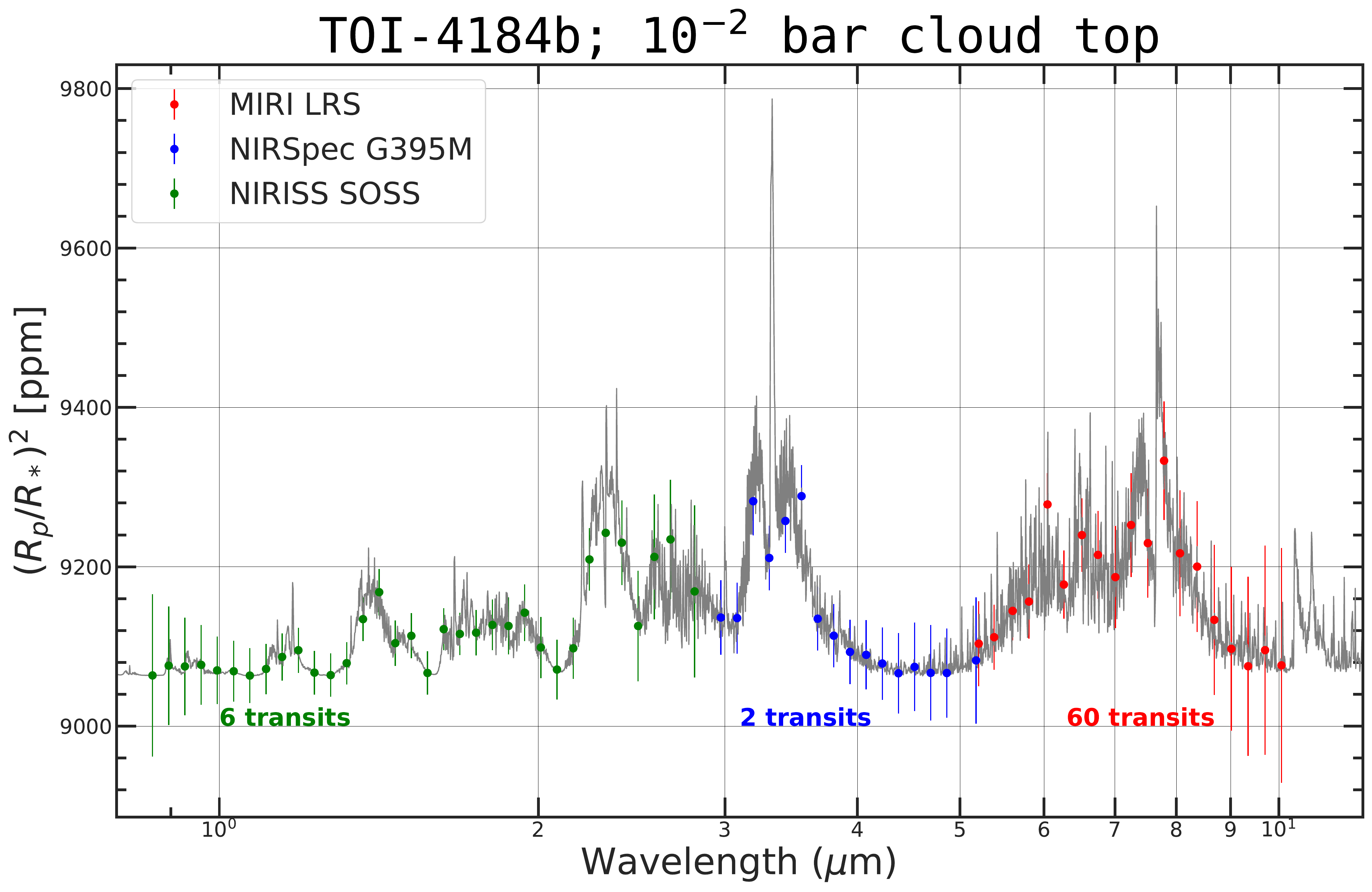} 
\caption{Simulated transmission spectra of TOI-4184\,b using NIRISS, NIRSpec, and MIRI instrument modes. This calculation assumes a cloudy H$_2$-rich atmosphere with a cloud top at 10$^{-2}$ bar. The initially modeled transmission spectrum is plotted in grey. The spectra have been binned to a resolution of R = 15.}
	\label{pan}		
\end{figure}
TOI-2084 and TOI-4184 are faint enough to be observable with all of JWST’s instruments. Using the JWST ETC PandExo \citep{bat}, we evaluated the detectability of the atmospheres of TOI-2084b and TOI-4184b with NIRISS-SOSS (0.6–2.8 $\mu$m), NIRSpec-G395M (2.88–5.20 $\mu$m) and MIRI-LRS (5–12 $\mu$m) instrumental modes for the clear and 10$^{-2}$ bar cloudy H$_2$-rich and water-rich atmospheric scenarios.  \autoref{pan} shows an example result for a cloudy H$_2$-rich atmosphere with a cloud top at 10$^{-2}$ bar for TOI-4184b. We find that NIRSpec and NIRISS observations are the most promising for this range of simulated atmospheres. A single transit observation with NIRSpec-G395M mode would be sufficient for robust detection of the molecular features in the clear H$_2$-rich scenarios for both planets,  whereas NIRISS-SOSS would require 2 transits. Two to three transits with NIRSpec-G395M could indeed characterize the atmospheres of TOI-2084 and TOI-4184, respectively, if they have clouds at 10$^{-2}$ bar. The spectral coverage of transit spectra could be extended to NIR wavelengths with NIRISS-SOSS observations, but this would require three to four times more transits to give a similar signal-to-noise ratio as of NIRSpec-G395M.\\
In summary, our simulations indicate that TOI-2084\,b and TOI-4184\,b will be great assets for exploring the nature of the atmospheres of sub-Neptunian exoplanets with JWST and upcoming next-generation space telescopes.


\begin{table*}
	\begin{center}
		{\renewcommand{\arraystretch}{1.17}
				\resizebox{0.9\textwidth}{!}{
			\begin{tabular}{llcccc}
				\hline
				Parameter & Symbol & Value              & Unit  \\
				          &        &   & \\
				\hline
				\textit{TOI-2084}      &  \\
				\hline
				Mean density    &  $\rho_\star$    &  $4.23^{+0.17}_{-0.16} $  &      $\rho_\odot$        \\
				Stellar mass    &  $M_\star$       & $0.453 ^{+0.029}_{-0.027}$   &      $M_\odot$            \\
				Stellar radius  &  $R_\star$       & $0.475^{+0.014}_{-0.014}$   &      $R_\odot$   \\
				Luminosity      & $L_\star$        & $0.0322 _{-0.0028}^{+0.0027}$        &  $L_\odot$    \\
                Effective temperature & $T_{\rm eff}$ & $3551^{+49}_{-52} $ & K \\ 
				Quadratic LD    & $u_{\rm 1,{\rm Sloan}-z'}$ &  $0.15 \pm 0.09$   &    \\
				Quadratic LD    & $u_{\rm 2,{\rm Sloan}-z'}$ &  $0.39 \pm 0.08$   &    \\
				Quadratic LD    & $u_{\rm 1,{\rm Sloan}-i'}$ &  $0.35 \pm 0.12$   &    \\
				Quadratic LD    & $u_{\rm 2,{\rm Sloan}-i'}$ &  $0.35 \pm 0.08$   &    \\
				Quadratic LD    & $u_{\rm 1,{\rm Sloan}-r'}$ &  $0.39 \pm 0.11$  &    \\
				Quadratic LD    & $u_{\rm 2,{\rm Sloan}-r'}$ &  $0.33 \pm 0.08$  &    \\
				Quadratic LD    & $u_{\rm 1,{\rm Sloan}-g'}$ &  $0.44 \pm 0.11$  &    \\
				Quadratic LD    & $u_{\rm 2,{\rm Sloan}-g'}$ &  $0.35 \pm 0.09$   &    \\
				Quadratic LD    & $u_{I+z'}$ &  $0.26 \pm 0.06$   &    \\
				Quadratic LD    & $u_{I+z'}$ &  $0.36 \pm 0.05$   &    \\
				Quadratic LD    & $u_{\rm 1,TESS}$ &  $0.21 \pm 0.10$   &    \\
				Quadratic LD    & $u_{\rm 2,TESS}$ &  $0.38 \pm 0.07$  &    \\
				\hline
				\textit{TOI-2084 b}      &  \\
				\hline
				Planet/star area ratio &  $(R_p/R_\star)^2_{{\rm Sloan}-z'}$ &  $ 2267^{+186}_{-179} $   &    ppm      \\
				                       & $(R_p/R_\star)^2_{{\rm Sloan}-i'}$ &  $1928^{+248}_{-241}$   &    ppm      \\
				                       & $(R_p/R_\star)^2_{{\rm Sloan}-r'}$ &  $1886^{+295}_{-327}$    &     ppm      \\
				                       & $(R_p/R_\star)^2_{{\rm Sloan}-g'}$  &  $2138^{+493}_{-442}$   &     ppm      \\
				                       & $(R_p/R_\star)^2_{I+z'}$       &   $2199^{+489}_{-454}$   &      ppm      \\
				                       & $(R_p/R_\star)^2_{\rm TESS}$     &  $1950^{+271}_{-273}$   &    ppm      \\  
				Impact parameter & $b' = a\cos i_p / R_\star$         &   $0.336 \pm 0.061$   &     $R_\star$       \\
				Transit duration &  $W$                               &   $122 \pm 2$    &     min \\
				Transit-timing & $T_0$                               &    $2458741.07323 \pm 0.00065$   &     BJD$_{\rm TDB}$     \\
				Orbital period  & $P$                              &     $6.0784247 \pm 0.0000096$  &   days     \\ 
				Scaled semi-major axis & $a_p/R_\star$    &  $22.66 \pm 0.39$   &       -   \\
				Orbital semi-major axis  & $a_p$          &  $0.05006 \pm 0.00103$   &       AU       \\
				Orbital inclination  &  $i_p$           &  $89.15 \pm 0.15$   &       deg    \\
				Radius          &  $R_p$                &  $2.47^{+0.13}_{-0.13}$   &       $R_\oplus $   \\
				Equilibrium temperature & $T_{\rm eq}$  &  $527 \pm 8$   &        K           \\
				Irradiation  & $S_p$                   &  $12.8 \pm 0.8$   &          $S_\oplus$  \\
				Predicted Mass               & $M_p$              &  $6.74^{+5.31}_{-2.81}$   &       $M_\oplus$           \\
				Predicted RV semi-amplitude         & $K$  &  $3.77^{+2.97}_{-1.57}$   &       $m/s$           \\
				\hline
		\end{tabular}}}
	\end{center}
	\caption{The TOI-2084 system parameters derived from our global MCMC analysis  (medians and $1\sigma$).}
	\label{toi2084_param}
\end{table*}


\begin{table*}
	\begin{center}
		{\renewcommand{\arraystretch}{1.14}
				\resizebox{0.9\textwidth}{!}{
			\begin{tabular}{llcccc}
				\hline
				Parameter & Symbol & Value              & Unit  \\
				          &        &   & \\
				\hline
				\textit{TOI-4184}      &  \\
				\hline
				Mean density    &  $\rho_\star$    &  $16.32^{+1.25}_{-1.09} $  &      $\rho_\odot$        \\
				Stellar mass    &  $M_\star$       & $0.2109 ^{+0.029}_{-0.026}$   &      $M_\odot$            \\
				Stellar radius  &  $R_\star$       & $0.2347^{+0.015}_{-0.015}$   &      $R_\odot$   \\
				Luminosity      & $L_\star$        & $0.00544 _{-0.00074}^{+0.00082}$        &  $L_\odot$    \\
                Effective temperature & $T_{\rm eff}$ & $3238^{+48}_{-49} $ & K \\ 
				Quadratic LD    & $u_{\rm 1,TESS}$ &  $0.21 \pm 0.029$   &    \\
				Quadratic LD    & $u_{\rm 2,TESS}$ &  $0.42 \pm 0.021$  &    \\
				Quadratic LD    & $u_{\rm 1,{\rm Sloan}-z'}$ &  $0.17 \pm 0.04$   &    \\
				Quadratic LD    & $u_{\rm 2,{\rm Sloan}-z'}$ &  $0.45 \pm 0.02$   &    \\
				Quadratic LD    & $u_{\rm 1,{\rm Sloan}-i'}$ &  $0.32 \pm 0.04$   &    \\
				Quadratic LD    & $u_{\rm 2,{\rm Sloan}-i'}$ &  $0.35 \pm 0.04$   &    \\
				Quadratic LD    & $u_{\rm 1,{\rm Sloan}-g'}$ &  $0.41 \pm 0.03$  &    \\
				Quadratic LD    & $u_{\rm 2,{\rm Sloan}-g'}$ &  $0.39 \pm 0.02$   &    \\
				Quadratic LD    & $u_{I+z'}$ &  $0.23 \pm 0.02$   &    \\
				Quadratic LD    & $u_{I+z'}$ &  $0.41 \pm 0.01$   &    \\
				Quadratic LD    & $u_{\rm 1,Ic}$ &  $0.26 \pm 0.05$  &    \\
				Quadratic LD    & $u_{\rm 2,Ic}$ &  $0.37 \pm 0.04$  &    \\
				Quadratic LD    & $u_{\rm 1,1.21\mu m}$ &  $0.02 \pm 0.01$  &    \\
				Quadratic LD    & $u_{\rm 2,1.21\mu m}$ &  $0.37 \pm 0.01$  &    \\
				\hline
				\textit{TOI-4184 b}      &  \\
				\hline
				Planet/star area ratio & $(R_p/R_\star)^2_{\rm TESS}$     &  $ 9115^{+937}_{-1000}$   &    ppm      \\ 
				                       &  $(R_p/R_\star)^2_{{\rm Sloan}-z'}$ &  $ 9491^{+560}_{-566} $   &    ppm      \\
				                       & $(R_p/R_\star)^2_{{\rm Sloan}-i'}$ &  $ 8461^{+540}_{-509}$   &    ppm      \\
				                       & $(R_p/R_\star)^2_{{\rm Sloan}-g'}$ &  $10982^{+3400}_{-2800}$    &     ppm      \\
				                       & $(R_p/R_\star)^2_{I+z'}$       &   $ 7869^{+970}_{-951}$   &      ppm      \\
				                       & $(R_p/R_\star)^2_{Ic}$       &   $ 8141^{+840}_{-770}$   &      ppm      \\
				                       & $(R_p/R_\star)^2_{1.21\mu m}$  &  $ 9334^{+2600}_{-2000}$   &     ppm      \\
				                       
				Impact parameter & $b' = a\cos i_p / R_\star$         &   $0.306 \pm 0.061^{+0.069}_{-0.094}$   &     $R_\star$       \\
				Transit duration &  $W$                               &   $ 77 \pm 1$    &     min \\
				Transit-timing & $T_0$                               &    $2459483.65667 \pm 0.00023$   &     BJD$_{\rm TDB}$      \\
				Orbital period  & $P$                              &     $4.9019804 \pm 0.0000052$  &   days     \\ 
				Scaled semi-major axis & $a_p/R_\star$    &  $30.79 \pm 0.96$   &     -     \\
				Orbital semi-major axis  & $a_p$          &  $0.0336 \pm 0.0015$   &       AU       \\
				Orbital inclination  &  $i_p$           &  $89.43 \pm 0.16$   &       deg    \\
				Radius          &  $R_p$                &  $2.43^{+0.21}_{-0.21}$   &       $R_\oplus $   \\
				Equilibrium temperature & $T_{\rm eq}$  &  $412 \pm 8$   &        K           \\
				Irradiation  & $S_p$                   &  $4.8 \pm 0.4$   &          $S_\oplus$  \\
				Predicted Mass               & $M_p$              &  $ 6.60^{+5.20}_{-2.75}$   &       $M_\oplus$           \\
				Predicted RV semi-amplitude         & $K$  &  $6.38^{+5.03}_{-2.66}$   &       $m/s$           \\
				\hline
		\end{tabular}}}
	\end{center}
	\caption{The TOI-4184 system parameters derived from our global MCMC analysis  (medians and $1\sigma$).}
	\label{toi4184_param}
\end{table*}

\section{Acknowledgements}
This paper made use of data collected by the TESS mission and are publicly available from the Mikulski Archive for Space Tele- scopes (MAST) operated by the Space Telescope Science Institute (STScI). Funding for the TESS mission is provided by NASA’s Science Mission Directorate. We acknowledge the use of public TESS data from pipelines at the TESS Science Office and at the TESS Science Processing Operations Center. Resources supporting this work were provided by the NASA High-End Computing (HEC) Program through the NASA Advanced Supercomputing (NAS) Division at Ames Research Center for the production of the SPOC data products.
Some of the observations in the paper made use of the High-Resolution Imaging instruments ‘Alopeke and Zorro obtained under Gemini LLP Proposal Number: GN/S-2021A-LP-105. ‘Alopeke and Zorro were funded by the NASA Exoplanet Exploration Program and built at the NASA Ames Research Center by Steve B. Howell, Nic Scott, Elliott P. Horch, and Emmett Quigley. Alopeke/Zorro was mounted on the Gemini North/South) telescope of the international Gemini Observatory, a program of NSF’s OIR Lab, which is managed by the Association of Universities for Research in Astronomy (AURA) under a cooperative agreement with the National Science Foundation. on behalf of the Gemini partnership: the National Science Foundation (United States), National Research Council (Canada), Agencia Nacional de Investigación y Desarrollo (Chile), Ministerio de Ciencia, Tecnología e Innovación (Argentina), Ministério da Ciência, Tecnologia, Inovações e Comunicações (Brazil), and Korea Astronomy and Space Science Institute (Republic of Korea).
The research leading to these results has received funding from  the ARC grant for Concerted Research Actions, financed by the Wallonia-Brussels Federation. TRAPPIST is funded by the Belgian Fund for Scientific Research (Fond National de la Recherche Scientifique, FNRS) under the grant PDR T.0120.21. TRAPPIST-North is a project funded by the University of Liege (Belgium), in collaboration with Cadi Ayyad University of Marrakech (Morocco)
MG is F.R.S.-FNRS Research Director and EJ is F.R.S.-FNRS Senior Research Associate. L.D. is an F.R.S.-FNRS Postdoctoral Researcher.
The postdoctoral fellowship of KB is funded by F.R.S.-FNRS grant T.0109.20 and by the Francqui Foundation.
This publication benefits from the support of the French Community of Belgium in the context of the FRIA Doctoral Grant awarded to MT.
This research is in part funded by the European Union's Horizon 2020 research and innovation program (grants agreements n$^{\circ}$ 803193/BEBOP), and from the Science and Technology Facilities Council (STFC; grant n$^\circ$ ST/S00193X/1).
This work is based upon observations carried out at the Observatorio Astron\'omico Nacional on the Sierra de San Pedro M\'artir (OAN-SPM), Baja California, M\'exico. SAINT-EX observations and team were supported by the Swiss National Science Foundation (PP00P2-163967 and PP00P2-190080),  the Centre for Space and Habitability (CSH) of the University of Bern, the National Centre for Competence in Research PlanetS, supported by the Swiss National Science Foundation (SNSF), and UNAM PAPIIT-IG101321.
J.d.W. and MIT gratefully acknowledge financial support from the Heising-Simons Foundation, Dr. and Mrs. Colin Masson and Dr. Peter A. Gilman for Artemis, the first telescope of the SPECULOOS network situated in Tenerife, Spain.
C.M.\ acknowledges support from the US National Science Foundation grant SPG-1826583. Research at Lick Observatory is partially supported by a generous gift from Google.
B.V.R. thanks the Heising-Simons Foundation for support. 
F.J.P acknowledges financial support from the grant CEX2021-001131-S funded by MCIN/AEI/ 10.13039/501100011033.
T.~C.~H. received funding from the Europlanet 2024 Research Infrastructure (RI) programme under grant agreement No. 871149. The Europlanet 2024 RI provides free access to the world’s largest collection of planetary simulation and analysis facilities, data services and tools, a ground-based observational network and programme of community support activities.
This work is partly supported by JSPS KAKENHI Grant Number JP18H05439 and JST CREST Grant Number JPMJCR1761.
This paper is based on observations made with the MuSCAT3 instrument, developed by the Astrobiology Center and under financial supports by JSPS KAKENHI (JP18H05439) and JST PRESTO (JPMJPR1775), at Faulkes Telescope North on Maui, HI, operated by the Las Cumbres Observatory.
U.G.J. gratefully acknowledges support from the Novo Nordisk Foundation Interdisciplinary Synergy Program grant no. NNF19OC0057374  and from the European Union H2020-MSCA-ITN-2019 under grant No. 860470 (CHAMELEON).
This work makes use of observations from the LCOGT network. Part of the LCOGT telescope time was granted by NOIRLab through the Mid-Scale Innovations Program (MSIP). MSIP is funded by NSF.
This research has made use of the Exoplanet Follow-up Observation Program (ExoFOP; DOI: 10.26134/ExoFOP5) website, which is operated by the California Institute of Technology, under contract with the National Aeronautics and Space Administration under the Exoplanet Exploration Program.
KAC acknowledges support from the TESS mission via subaward s3449 from MIT.
We are grateful to the ESO/La Silla staff for their continuous support. We acknowledge funding from the European Research Council under the ERC Grant Agreement n. 337591-ExTrA. We thank the Swiss National Science Foundation (SNSF) and the Geneva University for their continuous support to our planet search programs. This work has been carried out within the framework of the National Centre of Competence in Research PlanetS supported by the Swiss National Science Foundation under grants 51NF40\_182901 and 51NF40\_205606. The authors acknowledge the financial support of the SNSF.
 Portions of this work were conducted 
at Lick Observatory, which was built on 
the unceded territory of the Ohlone (Costanoans), Tamyen and
Muwekma Ohlone tribes, whose people continue to maintain their political
sovereignty and cultural traditions as vital members of central and northern California communities.
\bibliographystyle{aa}
\bibliography{aa.bib}

\clearpage
\onecolumn
\begin{appendix}
\section{Posterior probability distribution for the TOI-2084 and TOI-4184 systems}

\begin{figure*}[hbt!]
	\centering
	\includegraphics[scale=0.35]{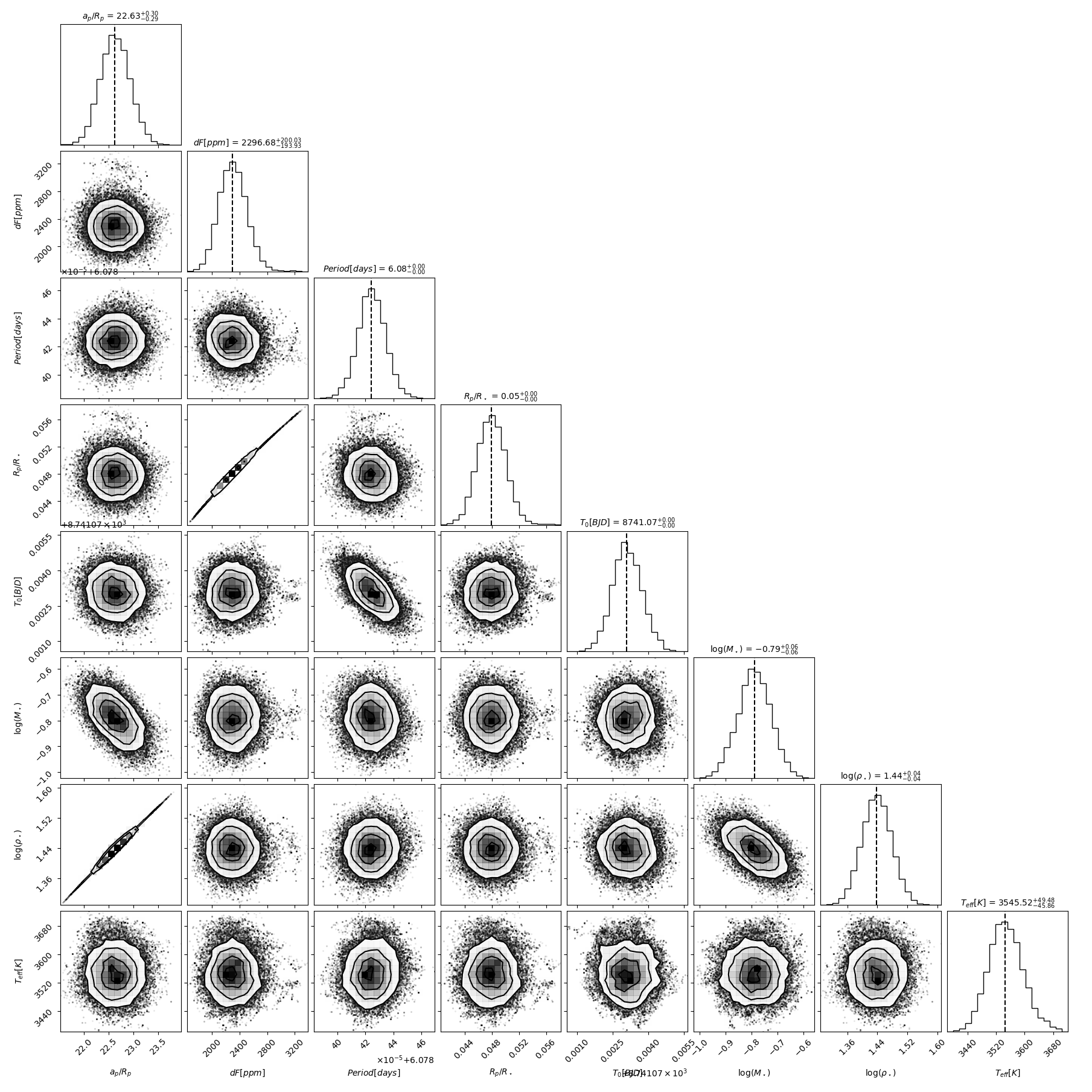}
	\caption{Posterior probability distribution for the TOI-2084 system stellar and planetary physical parameters fitted using our MCMC code as described in Methods. The vertical lines present the median value. The vertical dashed lines present the median value for each derived parameter.}
	\label{corner_TOI2084}
\end{figure*}

\begin{figure*}[hbt!]
	\centering
	\includegraphics[scale=0.35]{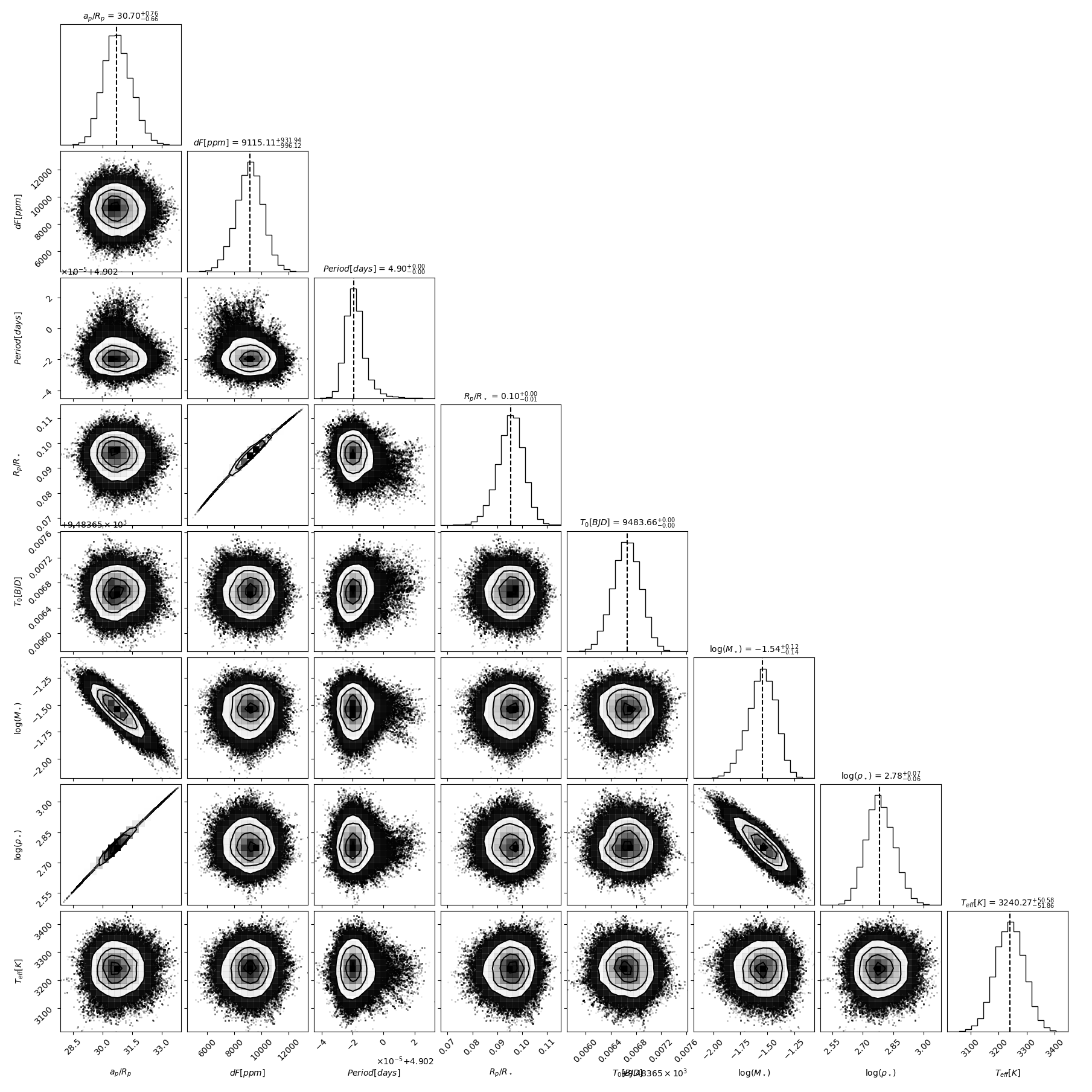}
	\caption{Posterior probability distribution for the TOI-4184 system stellar and planetary physical parameters fitted using our MCMC code as described in Methods. The vertical lines present the median value. The vertical dashed lines present the median value for each derived parameter.}
	\label{corner_TOI4184}
\end{figure*}

\end{appendix}

\end{document}